\documentclass[12pt,preprint]{aastex}
\usepackage{emulateapj5}

\newcommand{\cf}{{\ifmmode{C_f}\else{$C_{f}$}\fi}}
\newcommand{\zem}{{\ifmmode{z_{em}}\else{$z_{em}$}\fi}}
\newcommand{\zabs}{{\ifmmode{z_{abs}}\else{$z_{abs}$}\fi}}
\newcommand{\kms}{{\ifmmode{{\rm km~s}^{-1}}\else{km~s$^{-1}$}\fi}}
\newcommand{\deltrest}{{\ifmmode{\Delta t_{\rm rest}}\else{$\Delta t_{\rm rest}$}\fi}}
\newcommand{\voff}{{\ifmmode{v_{\rm shift}}\else{$v_{\rm shift}$}\fi}}
\newcommand{\cmm}{{\ifmmode{{\rm cm}^{-2}}\else{cm$^{-2}$}\fi}}
\newcommand{\cmmm}{{\ifmmode{{\rm cm}^{-3}}\else{cm$^{-3}$}\fi}}
\newcommand{\lya}{Ly$\alpha$} \newcommand{\Lya}{Ly$\alpha$}
\newcommand{\CIVdblt}{C~IV~$\lambda\lambda$1548,1551}
\newcommand{\NVdblt}{N~V~$\lambda\lambda$1239,1243}
\newcommand{\SiIVdblt}{Si~IV~$\lambda\lambda$1394,1403}
\newcommand{\ergpersec}{{\ifmmode{{\rm erg\;s}^{-1}}\else{${\rm erg\;s}^{-1}$}\fi}}
\newcommand{\Msolar}{{\ifmmode{{\rm M}_\odot}\else{${\rm M}_\odot$}\fi}}
\def\ten#1{{\times 10^{#1}}}
\newcommand{\ew}{{\ifmmode{{\rm EW}}\else{EW}\fi}}
\newcommand{\dew}{{\ifmmode{{\rm \Delta EW}}\else{$\Delta$EW}\fi}}
\newcommand{\ews}{{\ifmmode{{\rm EWs}}\else{EWs}\fi}}
\newcommand{\sn}{{\ifmmode{S/N}\else{$S/N$}\fi}}

\newcounter{species} 
\def\ion#1#2{\setcounter{species}{#2}#1$\;${\scriptsize\Roman{species}}\relax}

\slugcomment{\today}

\shorttitle{Monitor of NAL/mini-BAL Quasars}
\shortauthors{Misawa et al.}

\begin{document}

\title{Monitoring the Variability of Intrinsic Absorption Lines in 
  Quasar Spectra\altaffilmark{1,2,3}}

\footnotetext[1]{Based on data collected at Subaru Telescope, which is
  operated by the National Astronomical Observatory of Japan.}
\footnotetext[2]{Data presented herein were obtained at the W.M. Keck
  Observatory, which is operated as a scientific partnership among the
  California Institute of Technology, the University of California and
  the National Aeronautics and Space Administration. The Observatory
  was made possible by the generous financial support of the W.M. Keck
  Foundation.}  
\footnotetext[3]{Based on observations obtained at the European
  Southern Observatory at La Silla, Chile in programs 65.O-0063(B),
  65.O-0474(A), 67.A-0078(A), 68.A-0461(A), 69.A-0204(A),
  70.B-0522(A), 072.A-0346(A), 076.A-0860(A), 079.B-0469(A),
  166.A-0106(A)}

\author{Toru Misawa\altaffilmark{4}, 
        Jane C. Charlton\altaffilmark{5}, and
        Michael Eracleous\altaffilmark{5,6,7,8}}

\altaffiltext{4}{School of General Education, Shinshu University,
  3-1-1 Asahi, Matsumoto, Nagano 390-8621, Japan}
\altaffiltext{5}{Department of Astronomy \& Astrophysics, The
  Pennsylvania State University, University Park, PA 16802}
\altaffiltext{6}{Institute for Gravitation and the Cosmos, The
  Pennsylvania State University, University Park, PA 16802}
\altaffiltext{7}{Center for Relativistic Astrophysics, School of
  Physics, Georgia Institute of Technology, Atlanta, GA 30332}
\altaffiltext{8}{Department of Astronomy, University of Washington,
  Box 351580, Seattle, WA 98195}

\setcounter{footnote}{8}
\email{misawatr@shinshu-u.ac.jp}

\begin{abstract}
We have monitored 12 intrinsic narrow absorption lines (NALs) in five
quasars and seven mini-broad absorption lines (mini-BALs) in six
quasars for a period of 4--12 years (1--3.5 years in the quasar
rest-frame).  We present the observational data and the conclusions
that follow immediately from them, as a prelude to a more detailed
analysis. We found clear variability in the equivalent widths (\ews)
of the mini-BAL systems but no easily discernible changes in their
profiles. We did not detect any variability in the NAL systems nor in
narrow components that are often located at the center of mini-BAL
profiles.  Variations in mini-BAL \ews\ are larger at longer time
intervals, reminiscent of the trend seen in variable broad absorption
lines. If we assume that the observed variations result from changes
in the ionization state of the mini-BAL gas, we infer lower limits to
the gas density $\sim 10^3$--$10^5\;$\cmmm\ and upper limits on the
distance of the absorbers from the central engine of order a few
kpc. Motivated by the observed variability properties, we suggest that
mini-BALs can vary because of fluctuations of the ionizing continuum
or changes in partial coverage while NALs can vary primarily because
of changes in partial coverage.
\end{abstract}

\keywords{quasars: absorption lines -- quasars: individual
  (HE~0130-4021, HE~0940-1050, Q~1009+2956, HS~1700+6416,
  HS~1946+7658, Q~2343+125, UM~675, Q~1157+014, HE~1341-1020,
  HE~0151-4326, and HS~1603+3820)}

\section{INTRODUCTION\label{sec:intro}}
Quasars are routinely used as background sources, allowing us to study
the gaseous phases of intervening objects (e.g., galaxies, the
intergalactic medium, clouds in the halo of the Milky Way, and the
host galaxies of the quasars themselves) via absorption-line
diagnostics. A fraction of these absorption lines are produced by gas
ejected from the quasar central engine (i.e., an outflowing wind or
gas from the host galaxy that is swept up by this wind) rather than by
unrelated intervening structures. These absorbers that are {\it
  intrinsic} to the quasars themselves may be accelerated by
magnetocentrifugal forces \citep[e.g.,][]{eve05,dek95}, radiation
pressure in lines and continuum \citep{mur95,pro00}, radiation
pressure acting on dust \citep[e.g.,][]{voi93,kon94}, or a combination
of the above mechanisms.
The outflowing wind is an important component of a quasar central
engine: it can carry angular momentum away from the accretion disk,
facilitating accretion onto the black hole. Moreover, it may be the
origin of the broad emission lines that are the hallmark of quasar
spectra. In fact, the absorption-line and emission-line regions may be
different layers/phases of the wind
\citep[e.g.,][]{shi95,mur97,flohic12,chajet13}. Therefore, it is
necessary to understand the properties of the wind in order to
understand how quasar central engines work.
An outflowing wind also has the potential to influence the evolution
of the quasar host galaxy and its environs
\citep[e.g.,][]{silk98,king03} by (a) delivering energy and momentum
to the interstellar medium (ISM) of host galaxy and to the
intergalactic medium (IGM), thus inhibiting star formation
\citep[e.g.,][]{spr05}, and (b) transporting heavy elements from the
vicinity of the central engine to the IGM
\citep[e.g.,][]{ham97b,gab06}.  Therefore, to fully understand the
role that quasars play in galaxy evolution, we need to understand the
broader impact of ouflows from their central engines.

In view of the importance of quasar outflows, we have been focusing
our attention on rest-frame UV absorption lines from intrinsic
absorbers, in order to probe their demographics and physical
conditions. We have been primarily studying {\it narrow} absorption
lines (NALs; FWHM $\leq 500$~\kms) from intrinsic absorbers since
these allow us to probe wind properties in a unique way
\citep[e.g.,][]{ham97a,ham97b,mis07b}, unlike broad absorption lines
(BALs; FWHM $\geq$ 2,000 \kms; e.g., \citealt{wey91}), whose UV
resonance doublets are blended and may be saturated.  Equally useful
but more rare are mini-BALs, with intermediate widths between NALs and
BALs. Intrinsic NALs and mini-BALs present a powerful way to determine
the physical conditions of the outflowing gas. We are also interested
in finding out how the gas that gives rise to NALs and mini-BALs is
related to the gas producing the BALs.

The X-ray properties of quasars with intrinsic NALs suggest that the
corresponding absorbers are viewed along different lines of sight to
the central engine that those intercepting BAL gas
\citep{mis08,cha09,cha12,ham13}. In one possible picture emerging from
the models and observations, the BAL gas has the form of a fast,
dense, equatorial wind, following the models of \citet{mur95} and
\citet{pro00}, while the mini-BAL and NAL makes up the lower density
portion of the wind at high latitudes above the disk \citep[as
  suggested by][]{gan01}. In this context, NALs and mini-BALs are
valuable probes of different regions, different phases, and different
acceleration mechanisms of the wind. As noted by \citet{ham13} the
absence of strong X-ray absorption in mini-BALs suggests very strongly
that there is no thick shielding along those particular lines of sight
analogous to what is invoked for lines of sight along BALs (the
equivalent hydrogen column density in the shield can only be up to
$10^{22}\;{\rm cm}^{-2}$). In the absence of a thick shield, then,
implies that the intensity of the ionizing continuum must be lower so
as to avoid overionization of the gas and preserve the strong
\ion{Si}{4}, \ion{C}{4}, \ion{N}{5} that are commonly observed in
mini-BAL spectra.  An important implication of this constraint is that
the mini-BAL gas cannot be accelerated to high velocities (of order
0.1--0.3$c$) by radiative driving alone.

In an alternative picture, motivated by the models of \cite{kuro09},
the NAL gas may be gas from the host galaxy, at substantial distances
from the quasar central engine, swept up by an accretion disk
wind. This picture is bolstered by determinations of distances for the
NAL gas in a few quasars, that range from a few hundred pc to many kpc
from the source of the ionizing continuum \citep[e.g.,][]{ham01,
  ara13, borg12, borg13}. These determinations rely on constraints on
the density of the absorber, obtained from observations of metastable
absorption lines. If this is the appropriate picture, then NALs give
us a view of the evacuation of the gas from the host galaxy by a
quasar wind.

The variability of absorption lines is a very useful source of
information about the absorbing gas and can be used to test models for
the formation and structure of quasar winds. Thus, the variability of
BALs has been studied extensively. The probability that BALs will vary
over a given rest-frame time interval, \deltrest, has been found to
increase with \deltrest\ and approach unity for $\deltrest \gtrsim$ a
few years.  \citep[e.g.,][]{gib08,cap11,cap13,fil13}.  Variability is
usually seen only in portions of BAL troughs
\citep[e.g.,][]{cap11,cap12}, and detected more often in shallower
troughs at larger outflow velocities
\citep[e.g.,][]{lun07,gib08,fil13}.  In a relatively small number of
cases, BALs have transformed into mini-BALs by becoming narrower or
have disappeared completely
\citep[e.g.,][]{ham08,lei09,kro10,viv12,fil12}.  Possible origins of
this time variability include (a) motion of the absorbing gas parcels
across our line of sight \citep[e.g.,][]{ham08,gib08}, (b) changes in
the ionization state of the absorber \citep[e.g.,][]{ham11,mis07b},
and (c) redirection of photons around the parcels of gas that make up
the absorber by scattering material of variable optical depth,
resulting in time-variable dilution of the absorption troughs
\citep[e.g.,][]{lam04}. None of these mechanisms are applicable to
intervening absorbers because they have much larger sizes and lower
densities compared to intrinsic absorbers as discussed in
\citet{nar04}.

All of the above scenarios still appear to be plausible for BALs,
judging from their variability properties \citep{cap11,cap12,cap13}
and based on spectropolarimetric observations \citep[e.g.,
][]{goo95,lam04}, although \cite{fil14} favor scenario (b) based on
their results. In the case of mini-BALs scenarios (a) and (c) are
disfavored for at least one objects HS1603+3820.  Scenario (a) is
disfavored by the observation that all the troughs of the \ion{C}{4}
mini-BAL in this quasar vary in concert (i.e., the depths of all
troughs increase or decrease together) requiring an unlikely
coincidence between the motions of the individual parcels of gas
\citep{mis07b}.  Scenario (c) is disfavored because the polarization
level ($p$ $\sim$ 0.6\%) is not high enough throughout the spectrum to
reproduce the variation of absorption strength \citep{mis10} and, more
importantly, the polarization in the mini-BAL trough is the same as in
the continuum.
These results suggest that the most promising scenario for
variability, which may be applicable to a wide variety of absorbers,
is change of ionization conditions in the absorbing gas. Adopting this
hypothesis, and associating the observed variability time scale with
the recombination time in the absorbing gas, one can set a lower limit
to the electron density of this gas and an upper limit to its distance
from the source of ionizing radiation (see details in
\S\ref{sec:meth}). A number of authors have carried out this exercise,
obtaining lower limits on the electron density in the range
2,000--40,000~{\cmmm} and upper limits on the distance from the source
of ionizing radiation in the range 100--7,000~pc
\citep[e.g.,][]{ham97b,wis04,nar04,rod12}.
These values are consistent with those derived by applying
photoionization models to explain the strengths of absorption lines
from excited or metastable levels such as \ion{Fe}{2}, \ion{Si}{2}
\citep{kor08, dun10}, and \ion{O}{4} \citep{ara13}.  However, some of
these lines could be arising not in the outflowing wind but in the
interstellar medium of the host galaxies because their distances
($\sim$ several kpc) are quite large compared to the region of an
accretion disk from which an outflow may be launched ($\sim
10^{-3}$~pc for a $10^8\;\Msolar$ black hole) and the radius of the
broad-emission line region \citep[BLR; $\sim 0.2$~pc for a very
  luminous quasar of bolometric luminosity $\sim 4\times
  10^{47}\;\ergpersec$;][]{kas07}.

The physical mechanism for changing the ionization state of the
absorbing gas is not known.  An obvious choice would be variability of
the quasar UV continuum.  However, \citet{gib08} monitored 13 BAL
quasars over 3--6 years (in the quasar rest-frame) and found no
significant correlation between the change in flux density and the
strength of BALs.  Moreover, mini-BALs generally vary considerably
faster than the expected variability time scale of the UV continuum of
luminous quasars \citep[e.g.,][]{giv99,haw01,kas07}.  Nonetheless, the
fast variability of mini-BALs can be explained if the variations of
the ionizing continuum {\it seen by the absorber} is instead caused by
a porous/clumpy screen of variable optical depth between it and the
continuum source \citep{mis07b}.  This screen could be the inner part
of the outflow, perhaps a ``warm absorber'' such as those detected in
the X-ray spectra of BAL quasars \citep[e.g.,][]{gal02,gal06}.  Since
the ionization parameter and the total column density of warm
absorbers are at least two orders of magnitude higher than those of UV
absorbers \citep{ara01, ara13}, their variations should have a large
impact on the continuum that is transmitted through them and
illuminates UV absorbers.  Indeed, significant variations in the warm
absorbers and the soft X-ray flux that they transmit are detected
\citep{cha07,bal08,giu10a,giu10b}. This mechanism also allows for the
possibility that the {\it effective} size of the background source can
fluctuate, which will also lead to fluctuations in the apparent
\ews\ of the UV absorption lines due to a changing coverage fraction.

In this paper, we present results from a campaign to monitor the
variability of NALs and mini-BALs, building on previous such efforts
by other authors \citep{bar97,wam95,ham97b,wis04,nar04,mis05,hac13}.
We present the data and measurements of changes in the equivalent
widths (\ews) on time scales from months to years in the absorber's
rest frame.  We also discuss the conclusions that follow immediately
from these data. A more detailed analysis of coverage fractions,
profile variability, and comparison with photoionization models is
referred to future papers.  Our goals are to (a) constrain the
mechanism that causes the variability, and (b) place limits on the
density of the absorbers, which will then lead to limits on the
distance of the gas from the central engine. The distance of the gas
from the central engine, combined with other constraints, may also
allow us to distinguish between possible models for the NAL gas (e.g.,
filaments in the accretion-disk wind $vs.$ gas in the host galaxy that
has been swept up by the wind).
Thus, our campaign focuses on 12 NALs and 7 mini-BALs in 11 quasars
and employs high resolution spectra ($R > 36,000$), taken with 10-m
class telescopes (i.e., Subaru, Keck, and VLT), covering several
detected transitions.  At high resolution it is possible to resolve
internal narrower components in NALs and mini-BALs and monitor
variability of the line profiles, in addition to the variability of
the line strengths.  Ultimately, this information can be used to look
for the underlying cause of the variability and understand which of
the intrinsic absorption lines are due to the quasar wind, and which
are due to the circumgalactic medium of the host galaxy.

In \S2 we describe the target selection, observations, and data
reduction.  Our results concerning NAL and mini-BAL variability are
given in \S3.  In \S4 we discuss our results and in \S5 we summarize
our conclusions. We use a cosmology with H$_0$ = 72~\kms~Mpc$^{-1}$,
$\Omega_{\rm m}$ = 0.3, and $\Omega_{\Lambda}$ = 0.7.

\section{OBSERVATIONS AND DATA ANALYSIS\label{sec:obs}}

\subsection{Sample Selection}

We have selected quasars with intrinsic NALs based on the partial
coverage indicator: the dilution of absorption troughs by unocculted
light \citep[e.g.,][]{wam95,bar97} from the background sources. The
basic test relies on the optical depth ratio of resonant, rest-frame
UV doublets of Lithium-like species (e.g., \ion{C}{4},
\ion{N}{5}). When this ratio is inconsistent with 2:1, as dictated by
atomic physics (for fully resolved unsaturated absorption profiles),
the deviation is interpreted as dilution of the absorption troughs by
an unocculted portion of the background source. The optical depth
ratio provides a means of computing the fraction of background light
occulted by the absorber (the ``coverage fraction''): $C_f =
(R_1-1)^2/(R_2-2R_1+1)$, where $R_1$ and $R_2$ are the
continuum-normalized intensities of the weaker and stronger components
of the doublet. Recently, \citet{mis07a} surveyed 37 quasars at
redshift $2 < z < 4$ and found that at least 43\% had one or more
intrinsic NALs as evidenced by partial coverage in the \ion{C}{4},
\ion{N}{5}, and/or \ion{Si}{4} doublets, which implies that a
substantial fraction of the solid angle around a substantial fraction
of quasars is covered by material related to the quasar.  It is more
difficult to determine with certainty the origin of this material,
more specifically whether it originates in the host galaxy, or whether
it is a part of the quasar wind itself.

Our systematic study is based on long-term monitoring observations of
11 quasars (at $\zem\sim 2.00$--3.08) with at least one intrinsic NAL
or mini-BAL.  We select sample quasars from three sources; 1) quasars
with intrinsic NALs that were identified in \citet{mis07a}, 2) quasars
with mini-BALs that we identified from the VLT/UVES archive sample of
\citet{nar07}, and 3) quasars with mini-BALs already published in the
literature \citep{ham95,ham97b,dob99}.  These intrinsic NALs and
mini-BALs were identified through their partial coverage (as described
above) or their broad absorption profile.  Thus, our sample quasars
are all bright enough for high dispersion spectroscopy, and classified
into a category of very luminous quasars.  Another essential selection
criterion is that {\it at least two} previous high-signal-to-noise
ratio (\sn), high-resolution ($R=35,000$--45,000) spectra already
exist, taken at different epochs with either the Keck/HIRES, the
VLT/UVES and/or the Subaru/HDS.  The sample quasars that satisfy both
criteria are summarized in Table~\ref{tab:quasars}.

\subsection{Keck/HIRES Observations}
Four of our quasars come from a quasar sample that was originally
selected and observed for measuring the deuterium-to-hydrogen
abundance ratio (D/H) in the \Lya\ forest \citep[e.g.,][]{tyt96}. The
typical value of D/H is so small, 2--$4\times 10^{-5}$ \citep[][and
  references therein]{ome01}, that we can detect only \ion{D}{1} lines
corresponding to \ion{H}{1} lines with large column densities,
$\log$($N_{\rm H\;I}$/\cmm) $\geq 16.5$.  Therefore the survey
included 40 quasars, in which either damped \lya\ (DLA) systems or
Lyman limit systems (LLSs) were detected. This target selection method
does not directly bias our sample with respect to the properties of
any intrinsic absorption-line systems that these spectra may contain.
The observations were carried out by a group led by David Tytler,
using Keck/HIRES with a 1{\farcs}14 slit resulting in a velocity
resolution of $\sim 8$~\kms\ (FWHM). The spectra were extracted using
the automated program, MAKEE, written by Tom Barlow.
\citet{mis07a} used the spectra of 37 of these quasars, after removing
spectra of three quasars that cover only the \Lya\ forest region, and
identified 39 candidate intrinsic systems (28 reliable and 11 possible
systems) by partial coverage analysis. However, most spectra used in
\citet{mis07a} were produced by combining multi-epoch spectra in order
to increase the \sn. In this paper, we used 4 NAL quasars
(HE0130$-$4021, Q1009$+$2956, HS1700$+$6416, and HS1946$+$7658) out of
the 37 quasars above because their original spectra (i.e., before
combining data from multiple epochs) were available to us.
We also used a spectrum of the mini-BAL quasar (UM675) in one
epoch. The spectrum was obtained by Fred Hamann, using Keck/HIRES with
a 1{\farcs}15 \citep{ham97a}, who kindly provided it to us.

\subsection{VLT/UVES Observations}
Spectra of seven of the quasars studied here were obtained from the
VLT/UVES archive.  \citet{nar07} retrieved all R $\sim$ 45000 spectra
made available before 2006 June, and made a catalog of \ion{Mg}{2}
absorption systems in 81 quasar spectra. Since most of these quasars
were originally observed for studies of intervening absorbers, they
should not have a particular bias toward or against intrinsic NALs and
mini-BALs.  In this catalog, we found four mini-BAL systems in the
spectra of HE0151$-$4326, Q1157$+$014, HE1341$-$1020, and Q2343$+$125,
as well as intrinsic NAL systems in two quasars (HE0130$-$4021 and
HE0940$-$1050).  Some of the mini-BAL quasars above were already
discovered in past work \citep{ham97b,men01}.
In addition to these archival data, we re-observed three NAL quasars
(HE0130$-$4021, HE0940$-$1050, and Q1009$+$2956) and four mini-BAL
quasars (HE0151$-$4326, Q1157$+$014, HE1341$-$1020, and Q2343$+$125)
in 2007, using VLT/UVES with appropriate standard setups for each
target to cover the wavelength from {\lya} to
\ion{C}{4}~$\lambda\lambda$1548,1551 with only a few exceptions.  We
used a 1\farcs0 slit (yielding $R=40,000$; 7.5~\kms) and 2$\times$2
CCD binning. Neither the ADC nor the image-slicer were used.
We reduced the data following the procedure of \citet{nar07} using the
ESO provided MIDAS pipeline.  We also performed helio-centric velocity
correction and air-to-vacuum wavelength correction.

\subsection{Subaru/HDS Observations}
We obtained high-resolution spectra of one NAL quasar (HS1700$+$6416)
that was identified in \citet{mis07a} and two mini-BAL quasars (UM675
and HS1603$+$3820) from the literature \citep{ham97a,dob99} using
Subaru/HDS.  We chose these quasars because of their good visibility
from the Subaru telescope and their brightness, $m_V$ $\leq$ 17.  We
used a $1.\!\!^{\prime\prime}0$ slit (yielding R$\sim$36,000,
8.33km/s), the ADC, and the red-sensitive grating. The CCD binning was
set to 2$\times$1.  We used non-standard setups to cover \lya,
\NVdblt, \SiIVdblt, and \CIVdblt.
We reduced the data in a standard manner with the IRAF
software\footnote{IRAF is distributed by the National Optical
  Astronomy Observatories, which are operated by the Association of
  Universities for Research in Astronomy, Inc., under cooperative
  agreement with the National Science Foundation.}. Wavelength
calibration was performed using the spectrum of a Th-Ar lamp.
Because the blaze profile function of each echelle order changes with
time, we could not perform flux calibration using the spectrum of a
standard star. Therefore we directly fitted the continuum, which also
includes substantial contributions from broad emission lines, with a
third-order cubic spline function. Around heavily absorbed regions, in
which direct continuum fitting was difficult, we used the
interpolation technique introduced in \citet{mis03}. We have already
confirmed the validity of this technique by applying it to a stellar
spectrum.

\subsection{Data Summary and Spectroscopic Analysis}
To enhance the \sn\ of the spectra, all available observations of a
particular target, taken within 2 weeks (14 days) of each other, were
combined into a single spectrum, representing a single epoch. Although
some BAL features show variability on very short time scales
($\deltrest\leq 0.1$~years), the variation probability is much smaller
than that for longer time intervals \citep[e.g.,][]{cap13}.  Therefore
we prioritize increasing data quality above investigating short-term
variability (i.e., combining as many spectra as possible to increase
the \sn).
In total, we have monitored 12 intrinsic NALs in five quasars and 7
mini-BAL in six quasars for 4--12 years ($\sim1$--3.5 years in the
quasar rest-frame).  Each was observed with one or more of the
telescopes/spectrographs discussed above.  The observations are
summarized in Table~\ref{tab:obs}.

\begin{figure}
 \begin{center}
  \includegraphics[width=11cm,angle=0]{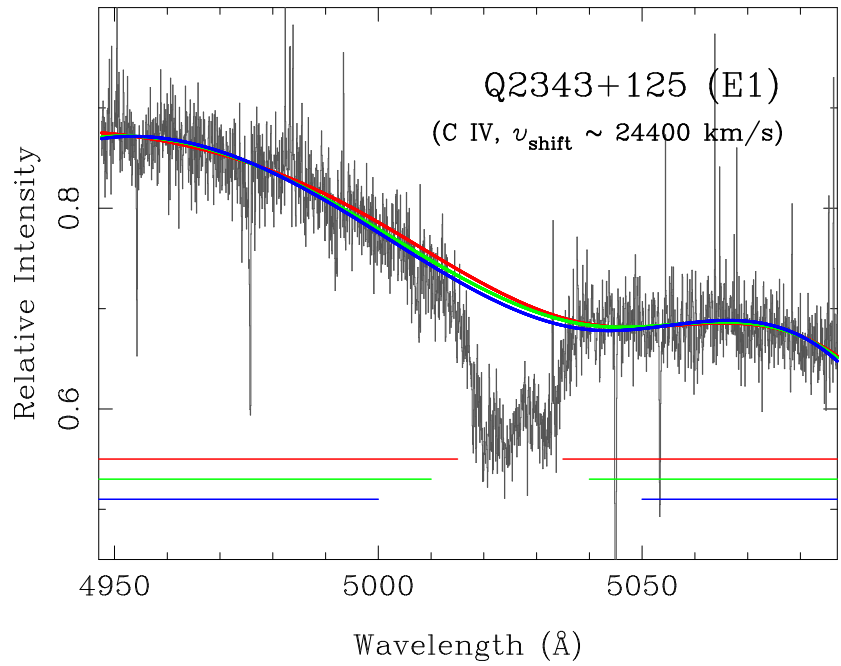}
 \end{center}
 \caption{An example of a continuum fit to the VLT/UVES spectrum (gray
   histogram) with a cubic spline function.  This spectrum covers the
   \ion{C}{4} mini-BAL at \voff\ $\sim$ 24,400~\kms\ in Q2343+125 and
   was obtained on Aug 18, 2002 (Epoch~1). Best-fitting continua are
   plotted with red, green, and blue curves whose fit regions are,
   5015 -- 5035~\AA, 5010 -- 5040~\AA, and 5000 -- 5050~\AA,
   respectively as shown with horizontal lines. Similar fits are
   applied for all VLT/UVES spectra, but the exact methodology differ
   for Keck/HIRES and Subaru/UVES spectra as discussed in
   \S\ref{sec:obs} of the text.\label{fig:fit_sample}}
\end{figure}

For the data analyses in the following sections, we normalized all
spectra as follows. We fit the VLT/UVES spectra (including both the
continuum and the broad emission lines) with cubic spline functions
and divided the original spectra by these functions. While fitting, we
avoid absorbed regions so that the fit interpolates across them. An
example of continuum fits around the \ion{C}{4} mini-BAL of the
VLT/UVES spectrum of Q2343+125 (Epoch~1) is shown in
Figure~\ref{fig:fit_sample}.  Because the continuum level usually
depends on the regions used for the fit, we experimented with three
different fit regions. The uncertainty from the continuum placement
will be discussed in \S\ref{sec:meth}. For Keck/HIRES and Subaru/HDS,
we separately fit each echelle order of the spectra with a cubic
spline function that includes the instrumental blaze function as well
as the continuum and the broad emission lines. This was necessary
because flux calibration was not applied to these spectra. (i.e., each
echelle order is not smoothly connected to adjacent orders.)

The normalized profiles of {\Lya}, {\NVdblt}, {\SiIVdblt}, and
{\CIVdblt}, superimposed for all epochs of observation (listed in
Table~\ref{tab:obs}), of the 12 intrinsic NALs and 7 mini-BALs are
displayed in Figures~\ref{fig:varplo}--\ref{fig:varplo_Q2343}, if they
are covered by the observed spectra. The different epochs are
represented by curves of different colors, with the 1st to 6th epochs
shown as black, red, green, blue, light-blue, and purple,
respectively. The velocity scale is relative to the flux-weighted
center of a line, calculated by equation~(4) of \citet{mis07a}.
For this display the mini-BAL spectra, only, were resampled every
0.3\AA\ because their \sn\ are relatively low and their profiles are
broad enough that no information is lost.  The {\Lya} and other metal
absorption lines of mini-BALs are omitted from this plot if they are
badly blended with the \lya\ forest and/or if they are observed with
Subaru/HDS, because in that case the continuum fit is too uncertain to
allow a variability analysis.

\section{MEASUREMENTS AND RESULTS}

\subsection{Methodology and General Results\label{sec:meth}}

Our initial inspection of the normalized spectra shown in
Figures~\ref{fig:varplo}--\ref{fig:varplo_Q2343} did not reveal any
substantial variations in the profiles of either the NALs or the
mini-BALs i.e., no changes in the profile asymmetries or the
velocities of the troughs were discernible. However, changes in the
strengths of the mini-BALs were quite evident. To quantify changes in
line strengths, we measured the \ews\ of the blue components of NALs
and the entire doublet regions of mini-BALs in each of the observed
epochs for each transition, i.e., \ion{C}{4}, \ion{Si}{4}, and
\ion{N}{5}. We list these measurements in Table~\ref{tab:ew} along
with their uncertainties.  For the mini-BALs, the doublet members
often suffer self-blending and cannot be reliably separated.  We
calculated the 1$\sigma$ error bar on the measured \ew, $\sigma_\ew$,
as the sum of contributions from uncertainties in the intensities of
individual pixels in the spectrum and from uncertainties in the
placement of the continuum:
$\sigma_\ew=\left(\sigma_{\rm pix}^2+\sigma_{\rm cont}^2\right)^{1/2}$.
The first term was evaluated by summing in quadrature the \ew\ error
bars from the $N$ individual pixels making up the absorption trough
(see horizontal arrows in
Figures~\ref{fig:varplo}--\ref{fig:varplo_Q2343})
\begin{equation}
  \sigma_{\rm pix}^2= 
  \sum_{i=1}^N \left(\sigma_i \, \Delta\lambda_i \right)^2.
\label{eq:sigpix}
\end{equation}
where $\sigma_{i}$ is the error in the normalized intensity at pixel
$i$, and $\Delta\lambda_i$ is the width of that pixel (in \AA). To
compute the uncertainty resulting from the continuum placement, we
assumed that it is proportional to the product of the width of the
line near the continuum level, $\lambda_{\rm max}-\lambda_{\rm min}$
(see horizontal arrows in
Figures~\ref{fig:varplo}--\ref{fig:varplo_Q2343}), and the
characteristic uncertainty of the continuum next to the line,
$\sigma_f$. In other words, we supposed that the fitted continuum
level in the spectrum could fluctuate by an amount of order $\pm
\sigma_f$, sweeping out an area of order $\pm (\lambda_{\rm
  max}-\lambda_{\rm min})\sigma_f$, which sets $\sigma_{\rm
  cont}$. The fluctuations in the {\it normalized} continuum level are
$\sigma_f/f$, where $f$ is the observed flux density, which is the
inverse of the signal-to-noise ratio, \sn. Hence we may write
\begin{equation}
  \sigma_{\rm cont}
  = A\, (\lambda_{\rm  max}-\lambda_{\rm min})\,{\sigma_f\over f}
  = {A\, (\lambda_{\rm  max}-\lambda_{\rm min})\over S/N}
\label{eq:sigcont}
\end{equation}
In the above expression we have included a ``calibration parameter,''
$A$, whose value we found empirically as follows. We carried out
several experiments in which we fitted the continuum around some NALs
and mini-BALs multiple times, choosing different regions around them
each time, so as to determine the amount by which the fitted level
could fluctuate. We then compared the variation in \ew\ measured after
different continuum fits to that given by equation~(\ref{eq:sigcont})
and found that $A\approx0.5$ very uniformly (see
Figure~\ref{fig:fit_sample}). Thus we adopted equation
~(\ref{eq:sigcont}) with $A=0.5$ as an estimate of the
\ew\ uncertainty resulting from continuum placement errors. We find
that the continuum placement errors are the dominant source of
uncertainty in the measured \ews.

To assess the significance of observed variations in the \ew\ we
computed the change in \ew\ between two epochs as $\Delta \ew = \ew_2
- \ew_1$ (where epoch 1 is earlier than epoch 2) and its uncertainty
as $\sigma_{\Delta \ew} =
\left({\sigma^2_{\ew1}+\sigma^2_{\ew2}}\right)^{1/2}$ and defined the
     {\it variation significance} as
\begin{equation}
  S_{\rm \ew}
  \equiv{\Delta \ew\over\sigma_{\Delta \ew}}
  ={\ew_2 - \ew_1 \over \left(\sigma^2_{\ew1}+\sigma^2_{\ew2}\right)^{1/2}}\; .
  \label{eq:sew}
\end{equation}

\begin{figure}
 \begin{center}
  \includegraphics[width=12cm,angle=0]{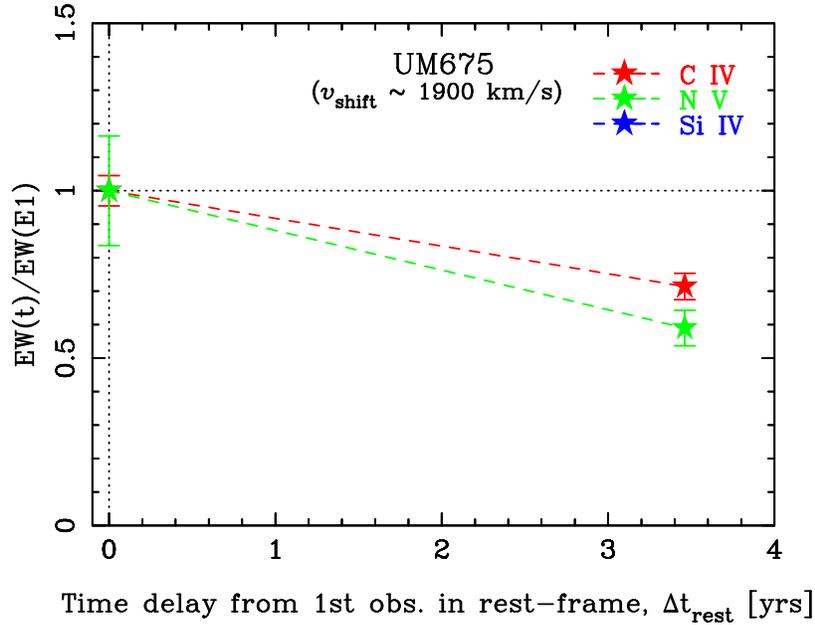}
 \end{center}
 \caption{Monitoring results for \ews\ of \ion{C}{4} (red), \ion{N}{5}
   (green), and \ion{Si}{4} (blue) lines in the mini-BAL at
   \voff\ $\sim$ 1,900~\kms\ in the UM675 spectra. The horizontal axis
   is the time delay from the first epoch in the quasar-rest frame (in
   years).  The vertical axis is the observed \ew\ (with 1$\sigma$
   errors) normalized to that of the first observed epoch. If the
   observed spectrum does not cover a particular line, that epoch is
   not plotted for that line.  If lines are covered but not detected,
   upper limits are marked with downward arrows.\label{fig:ew_um675}}
\end{figure}

\begin{figure}
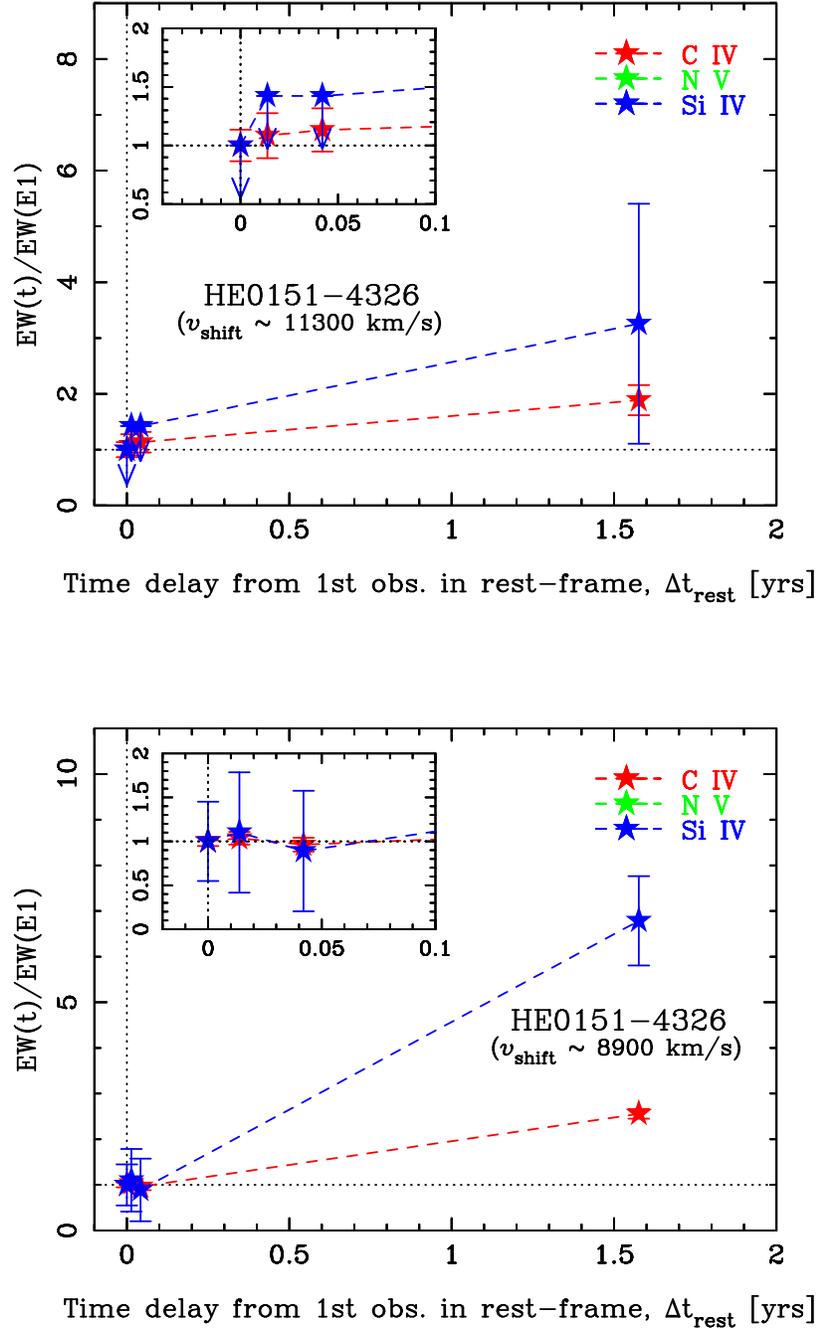

 \begin{center}
  \includegraphics[width=12cm,angle=0]{figure15a.eps}
  \includegraphics[width=12cm,angle=0]{figure15b.eps}
 \end{center}
 \caption{Same as Figure~\ref{fig:ew_um675}, but for a mini-BAL at
   \voff\ $\sim$ 11300~\kms\ (top) and 8900~\kms\ (bottom) in the
   spectra of HE0151$-$4326.\label{fig:ew_he0151_2.64}}
\end{figure}

\begin{figure}
 \begin{center}
  \includegraphics[width=12cm,angle=0]{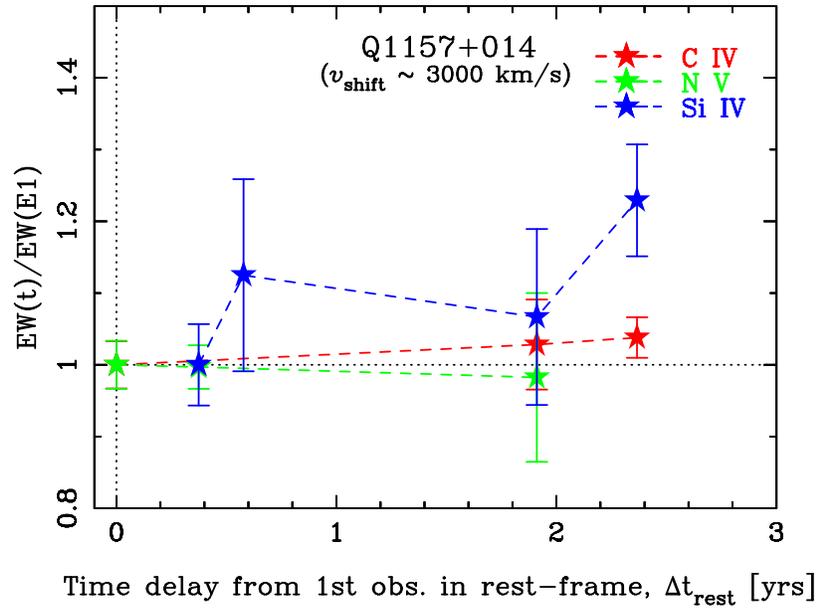}
 \end{center}
 \caption{Same as Figure~\ref{fig:ew_um675}, but for a mini-BAL at
   \voff\ $\sim$ 3000~\kms\ in the spectra of
   Q1157$+$014.\label{fig:ew_q1157}}
\end{figure}

\begin{figure}
 \begin{center}
  \includegraphics[width=12cm,angle=0]{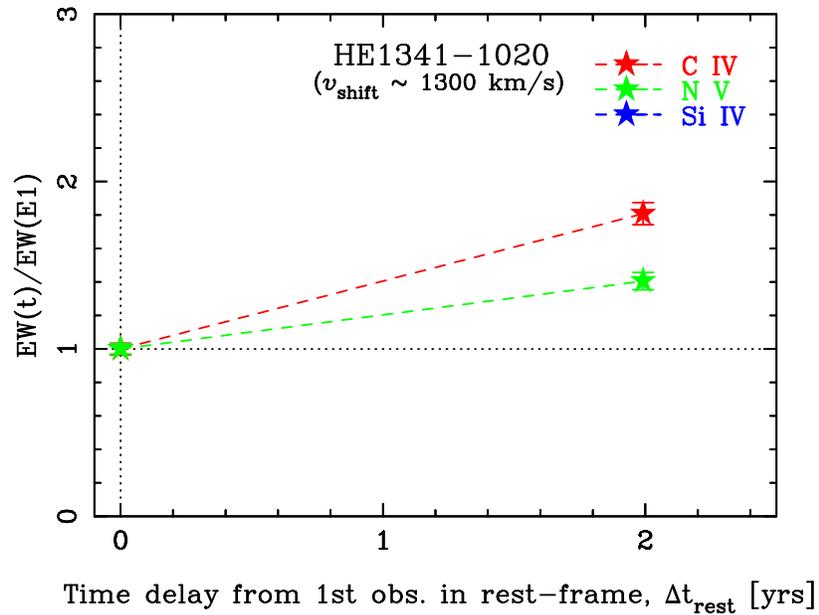}
 \end{center}
 \caption{Same as Figure~\ref{fig:ew_um675}, but for a mini-BAL at
   \voff\ $\sim$ 1300~\kms\ in the spectra of
   HE1341$-$1020.\label{fig:ew_he1341}}
\end{figure}

\begin{figure}
 \begin{center}
  \includegraphics[width=12cm,angle=0]{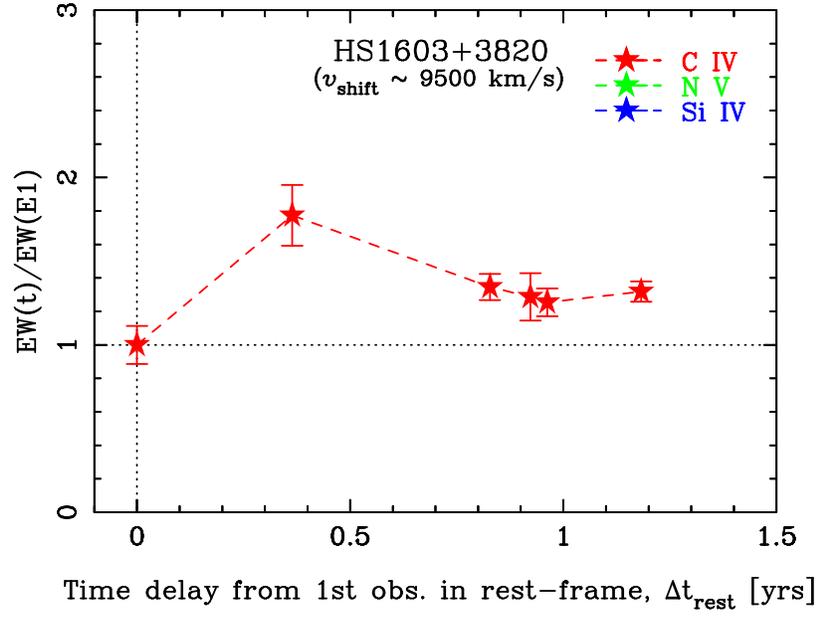}
 \end{center}
 \caption{Same as Figure~\ref{fig:ew_um675}, but for a mini-BAL at
   \voff\ $\sim$ 9500~\kms\ in the spectra of
   HS1603$+$3820.\label{fig:ew_hs1603}}
\end{figure}

\begin{figure}
 \begin{center}
  \includegraphics[width=12cm,angle=0]{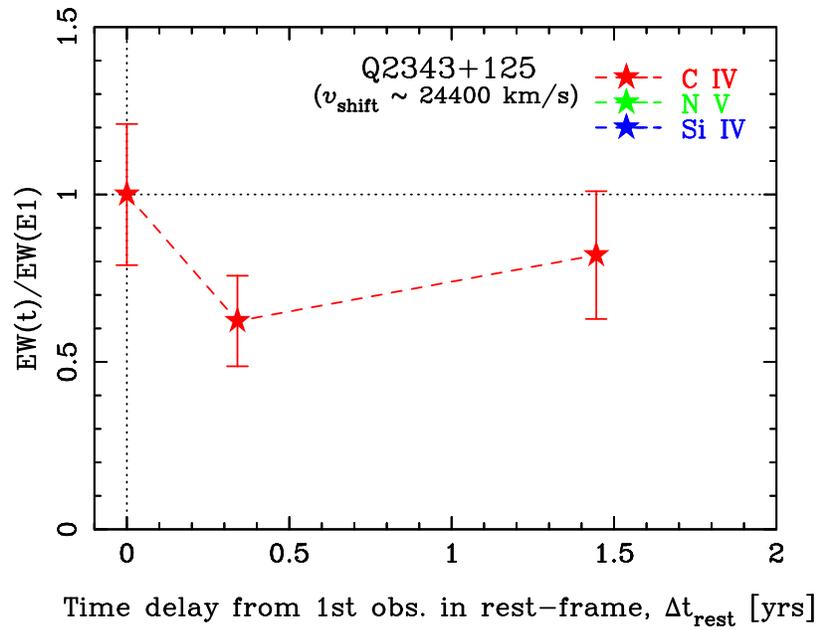}
 \end{center}
 \caption{Same as Figure~\ref{fig:ew_um675}, but for a mini-BAL at
   \voff\ $\sim$ 24,400~\kms\ in the spectra of
   Q2343$+$125.\label{fig:ew_q2343}}
\end{figure}

\begin{figure}
 \begin{center}
  \includegraphics[width=12cm,angle=0]{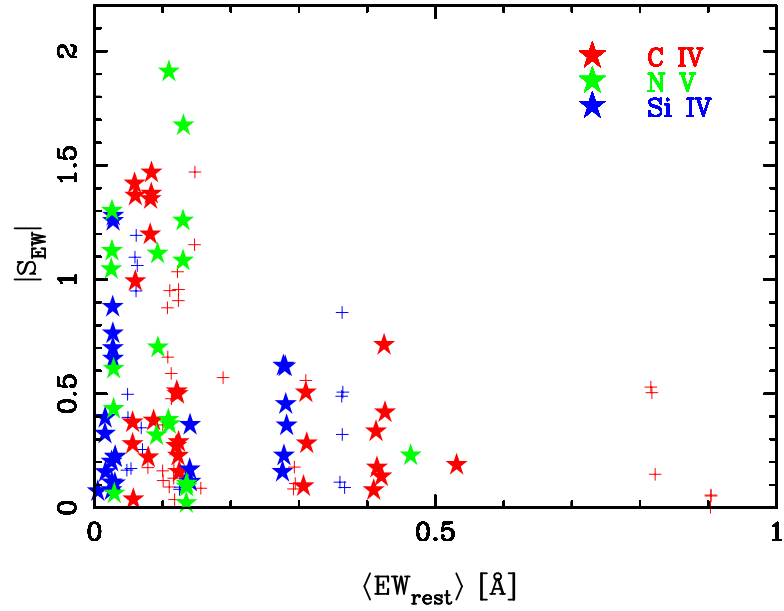}
 \end{center}
 \caption{Variation significance, $|S_\ew| \equiv |\dew|/\sigma_\dew$
   (see equation~\ref{eq:sew}), as a function of the average,
   rest-frame \ew\ for NALs. Intrinsic \ion{C}{4}, \ion{N}{5}, and
   \ion{Si}{4} NALs are marked with red, green, and blue stars, while
   intervening NALs with full coverage are marked with crosses and
   included for comparison.\label{fig:nalvary_ew}}
\end{figure}

\begin{figure}
 \begin{center}
  \includegraphics[width=12cm,angle=0]{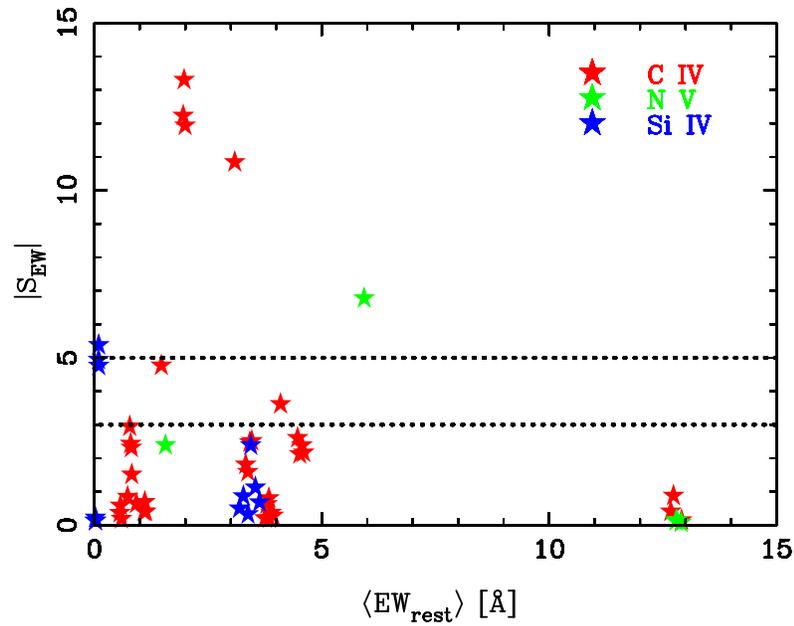}
 \end{center}
 \caption{Same as Figure~\ref{fig:nalvary_ew}, but for mini-BALs. The
   $|S_\ew|$ = 3 and 5 thresholds are marked with horizontal dotted
   lines.\label{fig:miniBALvary_ew}}
\end{figure}

\begin{figure}
 \begin{center}
  \includegraphics[width=12cm,angle=0]{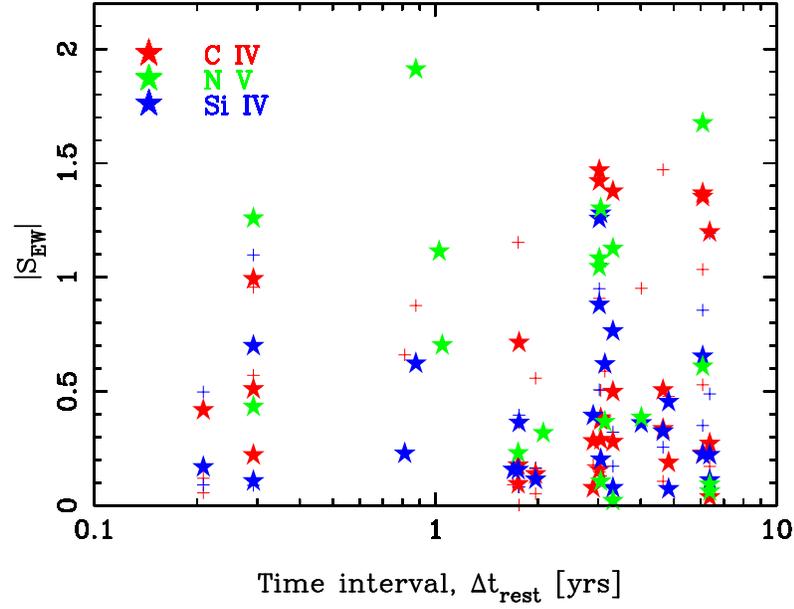}
 \end{center}
 \caption{Variation significance, $|S_\ew| \equiv |\dew|/\sigma_\dew$
   (see equation~\ref{eq:sew}), as a function of time interval for
   NALs. Intrinsic \ion{C}{4}, \ion{N}{5}, and \ion{Si}{4} NALs are
   marked with red, green, and blue stars, while intervening NALs with
   full coverage are marked with crosses for
   comparison.\label{fig:nalvary_dt}}
\end{figure}

\begin{figure}
 \begin{center}
  \includegraphics[width=12cm,angle=0]{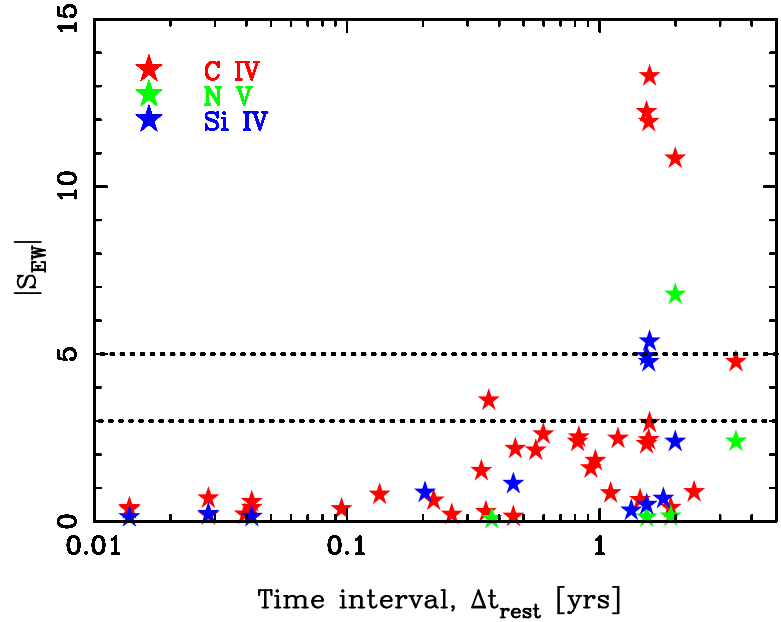}
 \end{center}
 \caption{Same as Figure~\ref{fig:nalvary_dt}, but for mini-BALs.  The
   $|S_\ew|$ = 3 and 5 thresholds are marked with horizontal dotted
   lines.\label{fig:miniBALvary_dt}}
\end{figure}

\begin{figure}
 \begin{center}
  \includegraphics[width=12cm,angle=0]{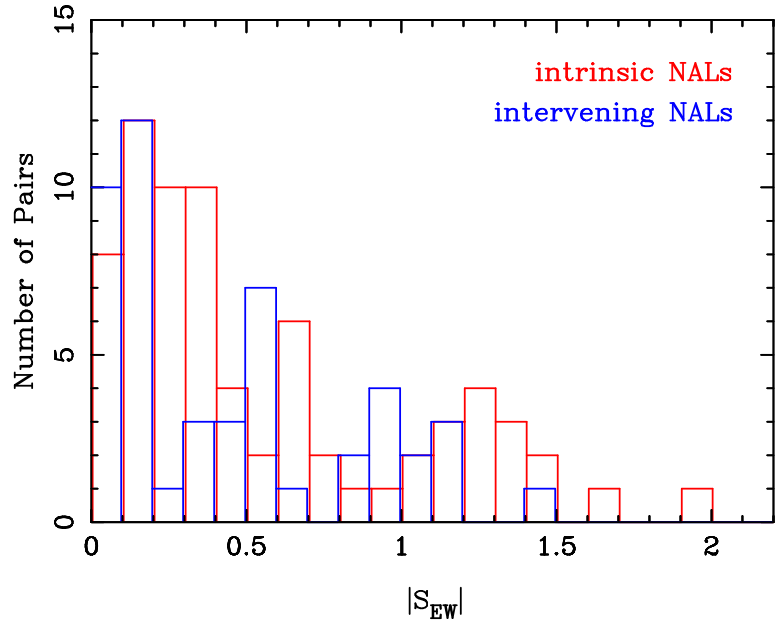}
 \end{center}
 \caption{The distribution of variation significance, $|S_{\ew}|\equiv
   |\dew|/\sigma_{\dew}$ (see equation~\ref{eq:sew}) for intrinsic
   NALs (red histogram) and intervening NALs (blue histogram).  We
   include all unique pairs of observing
   epochs.\label{fig:nalvary_hist}}
\end{figure}

\begin{figure}
 \begin{center}
  \includegraphics[width=12cm,angle=0]{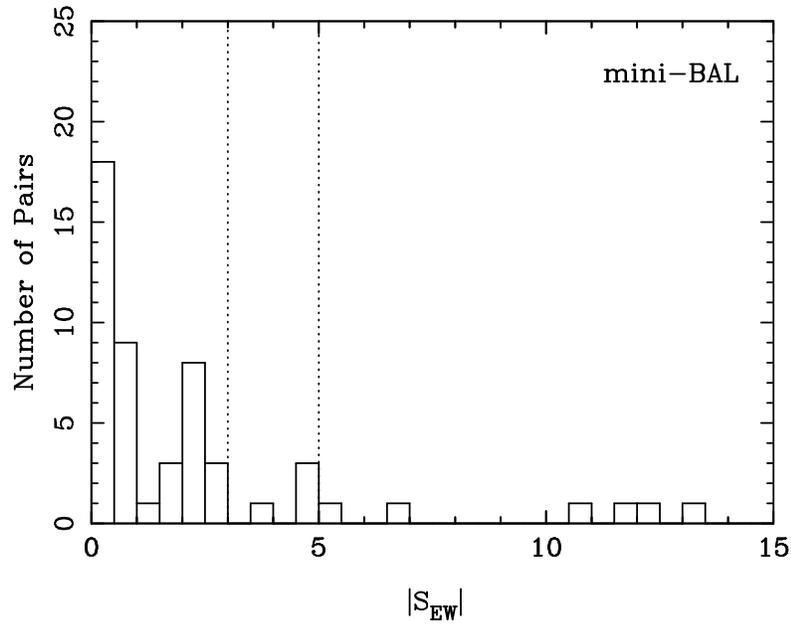}
 \end{center}
 \caption{Same as Figure~\ref{fig:nalvary_hist}, but for
   mini-BALs. The $|S_{\ew}|=3$ and 5 thresholds are marked with
   vertical dotted lines. \label{fig:miniBALvary_hist}}
\end{figure}

Significant variability is seen preferentially in the mini-BAL systems
(see Figures~\ref{fig:ew_um675}--\ref{fig:ew_q2343}), but not in NALs.
The greatest changes for NALs are seen from E1 $\rightarrow$ E2 of
\ion{N}{5} in the \voff\ $\sim$ 452~\kms\ system of Q1009+2956 and
from E2 $\rightarrow$ E4 of \ion{N}{5} in the \voff\ $\sim$
767~\kms\ system of HS1700+6416. In both of these cases the variation
significance is between 1.5$\sigma$ and 2$\sigma$, and this is
consistent with expectations in a sample of this size that does not
experience variability. It is also consistent with variations in the
\ews\ of intervening NALs caused by measurement errors.  We tabulate
the measured variation significance for mini-BALs in
Table~\ref{tab:prop} for the shortest time intervals where a variation
was detected (in the case of Q2343+125 we only detect variability at a
significance level of $< 2\sigma$ but we include this object in the
table for completeness). Distributions of absolute variation
significance, $|S_\ew|$, as a function of \ew\ and time interval are
shown in Figures~\ref{fig:nalvary_ew}--\ref{fig:miniBALvary_dt} for
NALs and mini-BALs, respectively.  Histograms of $|S_\ew|$, measured
for all unique pairs of observing epochs, are also shown in
Figures~\ref{fig:nalvary_hist} and \ref{fig:miniBALvary_hist}.

Five of the six mini-BAL systems in our sample show significant
variability at greater than $|S_\ew|>2$ in at least one transition (in
the sixth quasar, Q2343+125, the \ion{C}{4} line varied at
$|S_\ew|<2$; see Table~\ref{tab:prop}). Out of 52 unique pairs of
epochs plotted in the histogram of Figure~\ref{fig:miniBALvary_hist},
four are in the range $3<|S_\ew|<5$ while another 6 are in the range
$|S_\ew|>5$. For NALs, we compare our sample with intervening NALs
that are detected in the same spectra but are classified as
intervening NALs with coverage fractions of unity \citep{mis07a}.  In
Figure~\ref{fig:nalvary_hist} we compare the distribution of $|S_\ew|$
of intrinsic NALs to that of intervening NALs. This comparison shows
the two distributions to be very similar and reinforces our conclusion
that none of the intrinsic NALs in our sample have varied
significantly.

\begin{figure}
 \begin{center}
  \includegraphics[width=12cm,angle=0]{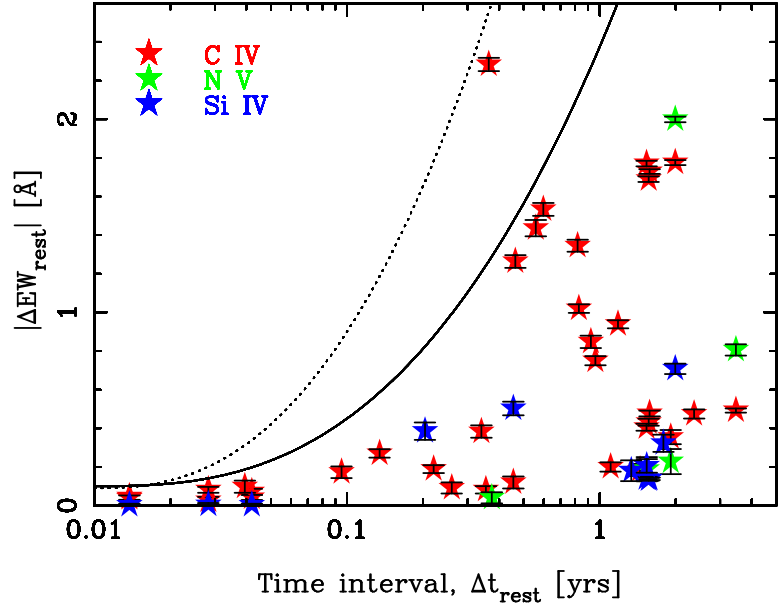}
 \end{center}
 \caption{Change in the absolute rest-frame \ew\ of mini-BALs as a
   function of rest-frame time interval for every unique pair of
   observing epochs. \ion{C}{4}, \ion{N}{5}, and \ion{Si}{4} data are
   shown with red, green, and blue stars with 1$\sigma$ errors,
   respectively.  The dashed and solid curves denote upper envelopes
   of the distribution with and without a outlying point at $|\Delta
   \ew_{\rm rest}|>2\;$\AA. They are included only as a guide to the
   eye. The outlying point corresponds to the system at
   \voff\ $\sim$9,500 \kms\ in HS1603$+$3820 (E1 $\rightarrow$ E2).
   Note that the horizontal axis is logarithmic.\label{fig:ewvary}}
\end{figure}

\begin{figure}
 \begin{center}
  \includegraphics[width=12cm,angle=0]{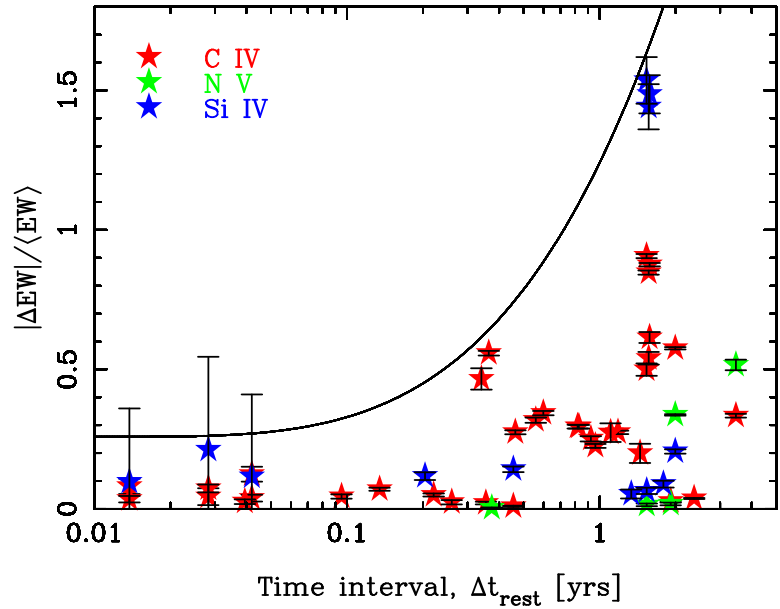}
 \end{center}
 \caption{Change in the fractional equivalent width ($|\Delta
   \ew|/\langle \ew\rangle$) as a function of rest-frame time
   interval. The vertical axis is the observed \ew\ (with 1$\sigma$
   errors) normalized to the average \ew\ in the observed-frame.  The
   solid curve denotes an upper envelope of the distribution and is
   included only as a guide to the eye. The symbols have the same
   meaning as in Figure~\ref{fig:ewvary} and the horizontal axis is
   logarithmic.\label{fig:ewvary_norm}}
\end{figure}

\begin{figure}
 \begin{center}
  \includegraphics[width=12cm,angle=0]{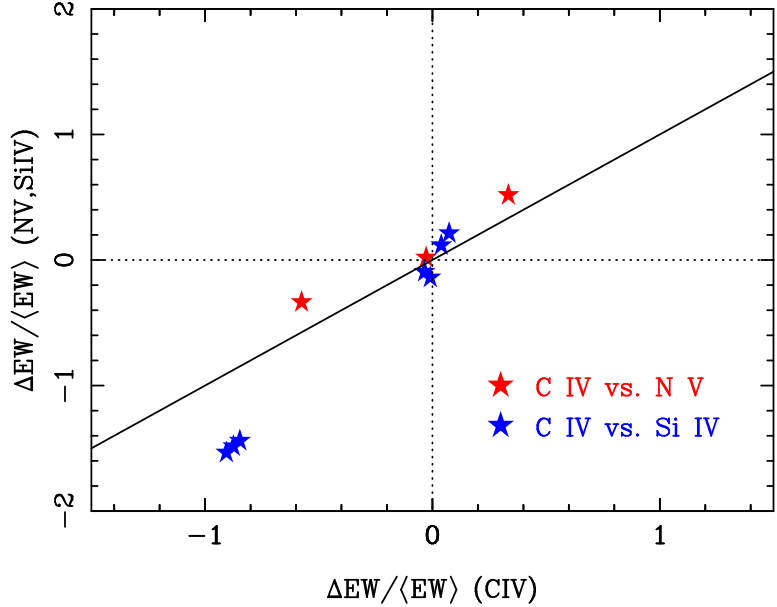}
 \end{center}
 \caption{Fractional variations of the observed \ews\ of the
   \ion{S}{4} and \ion{N}{5} mini-BALs plotted against the
   corresponding fractional variations of the \ion{C}{4} mini-BALs in
   the same systems. The \ew\ error bars are not plotted as they are
   smaller than the size of the symbols (see
   Figure~\ref{fig:ewvary}). In all but one case different lines from
   the same system vary together, in the same direction. The mini-BAL
   of Q1157+014, marked by the red star closest to the origin, appears
   to be an exception (the \ew\ of \ion{N}{5} appears to decrease
   while the other two lines appear to get stronger) but the
   statistical significance is very low. We also plot a straight solid
   line with unit slope for reference.
   \label{fig:ewvary_prop}}
\end{figure}

In Figure~\ref{fig:ewvary}, we plot the change in \ew\ of mini-BALs
for every unique pair of observing epochs, $|\dew|$, as a function of
the rest-frame time interval, \deltrest. The \ew\ changes are larger
for longer time intervals for lines from all observed ions,
\ion{C}{4}, \ion{N}{5}, and \ion{Si}{4}. The trend persists if we
normalize the \ew\ changes by the average \ew\ as shown in
Figure~\ref{fig:ewvary_norm}. In other words, both the absolute and
the fractional change in \ew\ increase over longer time intervals
(fractional variations of 50\% or more occur on time intervals of a
year or longer). Such a behavior has been noted for \ion{C}{4} BALs
\citep{gib08, cap11, cap13, fil13}. We find an analogous result
namely, when we detect significant variability in two transitions at
the same time in the same system the \ews\ change in the same
direction.  We illustrate this behavior in
Figure~\ref{fig:ewvary_prop} where we plot the change in the
\ion{Si}{4} or \ion{N}{5} \ew\ against that of \ion{C}{4} (see caption
for error bars and other specific information). It is noteworthy,
however, that we have never observed these three strong lines vary at
the same time because one of them is either saturated or not detected.
By comparing the data points in this figure with the straight line of
unit slope included for reference, we conclude that the fractional
variations of \ew\ of lines in the same system are not necessarily
proportional to each other.

If we assume that the observed changes in \ew\ result from changes in
the ionization state of the gas (see discussion in \S\ref{sec:intro}),
we can set a lower limit to the electron density by taking the
variability time scale, \deltrest, to be an upper limit to the
recombination time of the relevant ion, i.e., $n_e \gtrsim (\alpha\;
\deltrest)^{-1}$, where $n_e$ is the electron density and $\alpha$ is
the recombination coefficient of the transition in question. We use
the recombination coefficients for a gas temperature of 20,000~K
\citep{arn85,ham95}.  It is not straightforward to infer a density
from recombination to/from \ion{Si}{4} \citep{arn85}, therefore we
perform the calculations only for \ion{C}{4} and \ion{N}{5}. If the
\ews\ have increased, we assume that \ion{C}{5} $\rightarrow$
\ion{C}{4} and \ion{N}{6} $\rightarrow$ \ion{N}{5}, while if the
\ews\ have decreased, we assume that \ion{C}{4} $\rightarrow$
\ion{C}{3} and \ion{N}{5} $\rightarrow$ \ion{N}{4}, based on the
following argument. In all cases (with a single exception; see
previous paragraph) of variable mini-BAL systems absorption lines from
all ions (i.e., \ion{Si}{4}, \ion{C}{4}, and \ion{N}{5}, with
ionization potentials of 45.1, 64.5, and 97.9~eV, respectively) get
stronger and weaker together. In order for all three lines to increase
and decrease together, the ionization parameter should be either $\log
U \lesssim -2.5$ or $\log U \gtrsim -1.5$ \citep{ham97c}.  The regime
of low $U$ can be rejected because we do not detect any absorption
lines from low ionization species with ionization potentials between
14 and 33~eV corresponding to any mini-BAL systems (e.g.,
\ion{O}{1}~$\lambda$1302, \ion{Fe}{2}~$\lambda$2344,
\ion{Fe}{2}~$\lambda$2383, \ion{Si}{2}~$\lambda$1260,
\ion{Si}{2}~$\lambda$1527, \ion{Al}{2}~$\lambda$1671,
\ion{C}{2}~$\lambda$1335, \ion{Al}{3}~$\lambda$1855,
\ion{Al}{3}~$\lambda$1863, and \ion{Si}{3}).  This suggests that $\log
U \gtrsim -1.5$ and that as the ionization parameter varies, the main
ionization states of Silicon, Carbon, and Nitrogen vary as follows:
\ion{Si}{4} $\leftrightarrows$ \ion{Si}{5}, \ion{C}{4}
$\leftrightarrows$ \ion{C}{5}, and \ion{N}{5} $\leftrightarrows$
\ion{N}{6}. Under these conditions, we estimate lower limits on the
electron density in the range $1.7\times 10^3$--$4.0\times
10^5\;$\cmmm. We are only able to derive such limits for mini-BALs;
since the NALs in our sample did not show significant variability, we
are not able to apply this method to them.  We list the results for
individual objects in Table~\ref{tab:prop} and discuss them in more
detail in \S\ref{sec:details}

From the lower limits on the density of the mini-BAL gas we then
estimate upper limits to the distance of the absorber from the source
of ionizing radiation by making use of the definition of the
ionization parameter as the ratio of ionizing photon density to
hydrogen number density at the illuminated face of the cloud:
\begin{equation}
  U\equiv {n_{\gamma}\over n_H} = {Q(H)\over 4\pi r^2\, c\, n_H}\, ,
  \label{eq:U}
\end{equation}
where $Q(H)$ is the emission rate of hydrogen-ionizing photons by the
central engine and $r$ is the distance of the absorber from that
source.  We assume that $\log U > -1.5$ so that \ion{C}{4} is the
dominant ionization stage of Carbon \citep{ham95,ham97a}, and to
satisfy the constraint that all the observed absorption lines increase
and decrease together. We also assume that $n_e\sim n_H$.  Thus, we
obtain $Q(H)$ by scaling the bolometric luminosity, which we compute
via $L_{bol} = 4.4\,\lambda L_{\lambda}$ at $\lambda = 1450\;$\AA,
based on the SED of \citet{ric06}. This SED is a segmented power law
of the form $L_{\lambda}\propto \lambda^{\alpha}$ with $\alpha=-1.6$
from 1,000~\AA\ to 10~$\mu$m, $\alpha=-0.4$ from 10 to 1,000~\AA, and
$\alpha=-1.1$ from 0.1 to 10~\AA, following \citet{zhe97} and
\citet{tel02}. By integrating this SED, we find that
$Q(H)=10^{58.2}\;{\rm s}^{-1}$ for a quasar with a bolometric
luminosity of $10^{48}\;{\rm erg\;s}^{-1}$. Inserting the expression
for $Q(H)$ and limits on $n_e$ and $\log U$ into equation~(\ref{eq:U})
and rearranging we arrive at the following expression for the distance
of the absorber from the continuum source
\begin{equation}
  r \leq 1.18\;L_{\rm 48}^{1/2}\;n_5^{-1/2}\; {\rm kpc}\; ,
  \label{eq:rlim}
\end{equation}
where $L_{\rm 48}$ is the bolometric luminosity in units of
$10^{48}\;{\rm erg\;s}^{-1}$ and $n_5$ is the electron density in
units of $10^5\;{\rm cm}^{-3}$. Using equation~(\ref{eq:rlim}) we
estimate upper limits on the distance of the absorber to be in the
range 3--6~kpc (see Table~\ref{tab:prop} and \S\ref{sec:details}).
These values decrease by a factor of $\sim$4 if we use the SED of a
typical radio-loud quasar \citep{mat87} or a high luminosity
radio-quiet quasar \citep{dun10}.

\subsection{Notes on Individual Mini-BAL Systems\label{sec:details}}
Here we discuss the variable mini-BALs in more detail before moving on
to a general discussion about their properties and the possible causes
of variability.

\begin{description}

\item
UM675 ($\zem = 2.15$), mini-BAL system at $\voff\sim
1,900\;$\kms\ ($\zabs\sim 2.13$):

Time variability of the \ion{C}{4} and \ion{N}{5} doublets over
$\leq$2.9~yrs in the quasar rest frame was discovered in the mini-BAL
absorber at \zabs\ $\sim$ 2.13 (i.e., \voff\ $\sim$1900~\kms) in this
radio-quiet quasar \citep{ham95}.  Additional observations indicated
that the different ions can have different coverage fractions:
\cf\ $\sim$ 0.45 for \ion{C}{4}, \cf\ $\geq$ 0.83 for \lya, and
\cf\ $\geq$ 0.6 for \ion{O}{6} \citep{ham97a}.

We collected two spectra taken with Keck/HIRES and Subaru/HDS in 1994
and 2005, respectively. As shown in Figure~\ref{fig:varplo_UM675},
\ion{N}{5} and \ion{C}{4} got weaker at a $>$5$\sigma$ significance
over $\sim$3.5~years in the quasar rest frame.  \lya\ also showed the
same trend. The variation occurs across the entire profile, though
perhaps fractionally less in the central narrower component, as seen
in the spectrum of Q2343+125 \citep{ham97b}.  We cannot distinguish
between a change in ionization and a change in coverage fraction as
the cause of the variability.

{\it Summary:} The \ion{C}{4} and \ion{N}{5} profiles have weakened on
timescales of several years in the quasar rest frame.  This places
constraints on density ($n_e$ $\geq$ 3.3$\times$10$^3$ \cmmm), and
thus distance ($r$ $\leq$~4.0 kpc).  The detection of \ion{Ne}{8}
\citep{bea91} also suggests proximity to the quasar, although this
line is not covered by our optical spectra.

\item
HE0151$-$4326 ($\zem = 2.78$), mini-BAL system at $\voff\sim 11,300$
and 8,900~\kms\ ($\zabs\sim 2.64$ and 2.67):

This system has not yet been studied in detail as an intrinsic
absorption system.  There are two distinct absorption systems at
\voff\ $\sim$ 8900~\kms\ (corresponding to $\Delta v$ $\sim$ 0 --
1000~\kms\ in Figure~\ref{fig:varplo_Q0151}), with \ion{C}{4} and
\ion{Si}{4} detected, and at \voff\ $\sim$ 11300~\kms\ ($\Delta v$
$\sim$ $-$2800 to $-$1800~\kms), detected in \ion{C}{4}.  In addition
to these two systems, there is likely to be a weaker system at
\voff\ $\sim$ 6400~\kms\ ($\Delta v$ $\sim$ 1500 -- 3000~\kms) with
detected \ion{C}{4}.

This mini-BAL system spans more than 5000~\kms\ in velocity space,
although it does not satisfy the criterion for being a BAL (which
requires that the observed spectrum falls at least 10\% below the
model of continuum plus emission lines over a contiguous velocity
interval of at least 2000~\kms).

The first three epochs of observation are within one month, and
neither \ion{C}{4} nor \ion{Si}{4} showed variation in \ew\ at
$>$5$\sigma$ level, except for E2 $\rightarrow$ E3 of \ion{C}{4} for
the \voff\ $\sim$ 8900~\kms\ system. The last epoch is separated from
those first three by $\sim 2$~years in the quasar rest frame.  The
absorption is significantly stronger in the last epoch, again over the
entire velocity range of observation.  The \ion{Si}{4} is either not
detected or very weak in the first three epochs, but then has appeared
by the fourth epoch of observation in the component at $\Delta v$
$\sim$800~\kms\ (see Figure~\ref{fig:varplo_Q0151}).  This suggests
that there has been an ionization change in this component at least,
and not just a change in coverage fraction.  To adequately test this
hypothesis we would need high quality spectra covering some higher
ionization ions such as \ion{N}{5} and \ion{O}{6}.

{\it Summary:} The \ion{C}{4} profile for this absorption complex,
which spans $\sim$5000~{\kms}, has increased in strength between
observations separate by $\sim$2~years in the quasar rest frame.
\ion{Si}{4} has also appeared in one component.  This is consistent
with a lower degree of ionization at the later time, and places
constraints on densities ($n_e$ $\geq$ 3.9$\times$10$^3$ \cmmm), and
thus distances ($r$ $\leq$ 6.3~kpc) for the \voff\ $\sim$
11300~\kms\ and 8900~\kms\ systems.

\item
Q1157+014 ($\zem = 2.00$), mini-BAL system at $\voff\sim
3,000\;$\kms\ ($\zabs \sim 1.97$):

\citet{wri79} first studied the spectrum of this quasar in detail, and
noted that it has \ion{C}{4} and \ion{C}{3}] emission lines (although
  the \lya\ emission line is absent or extremely weak) and several
  broad absorption lines such as \lya, \ion{N}{5}, and \ion{C}{4}.
  This system is classified as a BAL by the traditional definition
  because of the broad absorption profiles of \ion{C}{4} and
  \ion{N}{5}. However, we regard it as a mini-BAL in this study
  because \ion{Si}{4} is not so strong as in typical BALs.

We collected 5 epochs of observations with VLT/UVES from 2000 -- 2007,
covering \ion{C}{4}, \ion{Si}{4}, and \ion{N}{5} at multiple epochs.
The profiles are shown in Figure~\ref{fig:varplo_Q1157}.  The
\ion{Si}{4} varies significantly on a timescale of $\leq$223~days
($\leq$80~days in the quasar rest frame).  The \ion{N}{5} appears to
get a bit weaker and the \ion{C}{4} to get a bit stronger over the
interval, but the change is at less than a 5$\sigma$ level except for
E1 $\rightarrow$ E4 for \ion{C}{4}.  The \ion{C}{4} and \ion{N}{5} are
clearly saturated over most of the profile, and so are not subject to
variations in column density.  A variation in coverage fraction would
lead to a significant change in those \ews, since non-unity coverage
fraction would cause these saturated profiles not to be black.  Thus
the lack of variability in the saturated \ion{C}{4} and \ion{N}{5}
suggests that an ionization change is responsible for the variability
of the \ion{Si}{4} profile in this system.

{\it Summary:} This system shows variability, but mostly in the weaker
\ion{Si}{4} absorption, and not in the saturated \ion{C}{4} and
\ion{N}{5}.  The most plausible explanation for this is that the
coverage fraction remains similar, but that the ionization state
changes, because a change in the \ion{C}{4} and \ion{N}{5} column
density won't make a significant difference in the profile if it is
saturated \citep[e.g.,][]{cap12}.  This would be consistent with these
profiles having changed only in their unsaturated parts.  Any strict
constraints cannot be placed on density and distance because
recombination from/to \ion{Si}{4} is not simple.

\item
HE1341$-$1020 ($\zem = 2.134$), mini-BAL system at $\voff\sim
1,300\;$\kms\ ($\zabs\sim 2.12$):

We found this mini-BAL system with broad absorption features of
\ion{C}{4} and \ion{N}{5} in the catalog of \citet{nar07}. As far as
we know, this system has not been previously studied in detail as an
intrinsic absorption system.

We collected two epochs of observations with VLT/UVES in 2001 and
2007, which are shown in Figure~\ref{fig:varplo_Q1341}.  With the two
available epochs we find that both lines show time variability on a
timescale of $\sim$2~yrs in the quasar rest frame.  Both the
\ion{C}{4} and the \ion{N}{5} are stronger at the later epoch at more
than 5$\sigma$ significance level, and they are stronger over the
entire profile.  \ion{Si}{4} is not detected at either epoch.  The
profiles again have the appearance of narrower profiles superimposed
on broad ones.

{\it Summary:} The \ion{C}{4} and \ion{N}{5} profiles have varied on a
timescale of $\sim 2$~years in the quasar rest frame, with both ions
becoming stronger over the full range in velocity.  This is certainly
consistent with a change in coverage fraction, but an ionization
change also cannot be ruled out with only two ions available as
constraints.  If an ionization change is the cause of variability, we
can place constraints on density ($n_e$ $\geq$ 3.0$\times$10$^3$
\cmmm), and distance ($r$ $\leq$ 3.5 kpc).  The absorption profiles
show narrow components superimposed on broader ones.

\item
HS1603+3820 ($\zem = 2.542$), mini-BAL system at $\voff\sim
9,500\;$\kms\ ($\zabs\sim 2.43$):

This system was first discovered by \citet{dob99}, and then monitored
with high-resolution spectra for $\sim$1.2~yrs in the quasar rest
frame \citep{mis05,mis07b}.  The \ion{C}{4} mini-BAL shows both
partial coverage \citep{mis03} and time variability
\citep{mis05,mis07b}.  The \ion{C}{4} profile changes across the
entire range of velocity, $\sim$2000~{\kms}, at once (see
Figure~\ref{fig:varplo_HS1603}).  Among three possible scenarios for
the variation: i) a gas motion, ii) a variable scattering material,
and iii) a change of ionization condition, the first two cannot be
applicable for this quasar (see \S1).  The preferred explanation of
changing ionization parameter is likely to be caused by variable
shielding material between the continuum source and the absorber,
since a quasar of this luminosity does not vary enough to explain the
change. Warm absorbers are a candidate for the shielding material as
evidenced by X-ray observations of this quasar \citep{mis10,roz13}.

{\it Summary:} There is significant variability of \ion{C}{4} over the
full velocity range on timescales as short as months in the quasar
rest frame.  The variability is likely to be caused by changing
ionization parameter due to variable shielding material that modulates
the quasar flux. We place constraints on the electron density ($n_e$
$\geq$ 1.6$\times$10$^4$ \cmmm) and the absorber's distance from the
flux source ($r$ $\leq$ 4.0 kpc)

\item
Q2343+125 ($\zem = 2.515$), mini-BAL system at $\voff\sim
24,400\;$\kms\ ($\zabs\sim 2.24$):

Previously, \citet{ham97b} detected \ion{C}{4}, \ion{N}{5}, and
\ion{Si}{4} mini-BALs at \zabs\ $\sim$ 2.24 (corresponding to
\voff\ $\sim$ 24400~\kms) in this radio-quiet quasar spectrum.  These
lines showed both time variability (the lines weakened and then
strengthened again by a factor of $\sim$4 on a timescale of
$\leq$0.3~yrs in the quasar rest frame) and partial coverage
(\cf\ $\leq$ 0.2 for \ion{C}{4}).

We collected additional spectra with VLT/UVES at three different
epochs from 2002 to 2007.  Because of the different wavelength
coverage of the different observations, only \ion{C}{4} has multiple
observations (Figure~\ref{fig:varplo_Q2343}).  \lya\ and \ion{N}{5}
are blended with the \lya\ forest.  The \ew\ measurements for
\ion{C}{4}~1548 were $3.29\pm 0.69$~\AA, $2.04\pm 0.44$~\AA, and
$2.69\pm 0.63$~\AA\ for observations 1, 2, and 3, respectively,
indicating only negligible variability at less than the 2$\sigma$
level. We do not detect any \ion{Si}{4} features, while \citet{ham97b}
found a weak \ion{Si}{4} mini-BAL.  Several narrow components inside
the broader absorption profiles of \ion{C}{4} \citep{ham97b} are not
apparent in our lower-\sn\ spectra.

{\it Summary:} In this \voff\ $\sim$ 24,400~\kms\ mini-BAL, \ion{C}{4}
gets slightly weaker, and then stronger, with changes taking place in
only 124 and 403 days in the quasar rest frame, but in $<$ 2$\sigma$
level.  More detailed analysis is not possible due to there being
multiple epoch observations only of \ion{C}{4}.

\end{description}

\section{DISCUSSION}

\subsection{Comparison of BAL and mini-BAL Variability and Assessment of Physical Mechanisms}

The observational results of our monitoring campaign, combined with
earlier case studies \citep[e.g.,][and references
  therein]{mis07b,ham11,rod12}, suggest some similarities between the
variability patterns observed in mini-BALs and BALs. In particular, we
have found that the \ews\ of all mini-BALs in our sample vary on
rest-frame time scales of months to years and have observed the
following patterns (a) when different transitions in a mini-BAL system
vary together, their \ews\ change in the same direction, (b) absolute
and relative changes in \ews\ are larger over longer rest-frame time
intervals, and (c) variations in the \ews\ of the lines are not
accompanied by discernible changes in the profiles, i.e., different
troughs in the profile vary together. In comparison (a) more than 50\%
of BALs vary over time scales of order a year in the quasar rest-frame
with the probability of variation approaching unity on multi-year time
scales \citep[e.g.,][]{cap11,cap13}, and (b) the \ews\ of the UV
absorption lines from \ion{C}{4}, \ion{Si}{4}, and \ion{N}{5} in BAL
systems vary together and in the same direction, although not
necessarily in proportion to each other \citep[e.g.,][]{fil13}.
However, BAL profiles do vary as the \ews\ vary
\citep[e.g.,][]{lun07,gib08,cap11,fil12,fil13} in contrast to what is
observed for mini-BAL troughs. In view of the similarities, we
consider the two variability mechanisms often discussed for BALs as
explanations for the observed mini-BAL variability, gas motion across
our sightline and change of ionization conditions (see
\S\ref{sec:intro} for references and more details)

To evaluate the likelihood that the mini-BAL variations result from
motion of the absorbing gas across the line of sight, we estimate the
time scale for a parcel of gas to cross the cylinder of sight that
encompasses either the continuum source or the BLR. We take the
transverse speed of the absorber to be of the same order as the
observed outflow speed (both are related to the free-fall speed, the
only velocity scale in the problem), i.e., $v\sim 10^4\;\kms$. We
assume that the observed UV continuum is emitted from the inner parts
of the accretion disk within a radius $a_{\rm cont} \sim 10\, r_{\rm
  g}$ from the center, where $r_{\rm g}\equiv GM_\bullet/c^2$ is the
gravitational radius and $M_\bullet$ is the mass of the black hole,
i.e., $a_{\rm cont} \sim 1.5\times 10^{15}\,M_9\;$cm, where $M_9 =
M_\bullet/10^9\,M_\odot$. Using the bolometric luminosities from
Table~\ref{tab:prop} and assuming that the accretion rates are close
to the Eddington limit, we estimate the black hole masses to be in the
range: $M_9=2$--15, which leads to continuum source radii in the range
$a_{\rm cont}\sim 3\times 10^{15}$--$2.3\times 10^{16}\;$cm and
crossing times $t_{\rm cross} (a_{\rm cont})\sim 0.09$--0.7~years. For
the radius of the BLR, we adopt the relation between it and the
continuum luminosity given by \citet{kas07}, which leads to $a_{\rm
  BLR}\sim 0.22$--0.63~pc and crossing times of $t_{\rm cross} (a_{\rm
  BLR})\sim 20$--60~years.  The long crossing times of the BLR lead us
to disfavor gas motion across the BLR cylinder of sight as the cause
of variability. Moreover, a substantial fraction of the mini-BAL
troughs in our sample are found at large outflow velocities, $|\voff|
\gtrsim 5,000\;\kms$, so that they do not overlap with the broad
emission lines (i.e., the mini-BAL gas absorbs only continuum photons
and the continuum source crossing time is the more relevant time
scale). These considerations make this particular version of the
scenario very unlikely. The crossing times of the continuum source for
the more luminous quasars in our sample are on a par with the observed
variability time scales and the amplitude of \ew\ variations is
largest for time scales just over a year, which is comparable to the
time required to traverse the continuum source completely. However,
for the less luminous quasars in our sample, we would have expected
the \ew\ variations to be faster. Moreover, all troughs in a mini-BAL
system vary together when the \ew\ varies, which is difficult to
orchestrate unless the parcels of gas producing the mini-BAL troughs
in a given system all cross the cylinder of sight to the continuum
source at the same time. Thus, we disfavor this version of the
scenario as the cause of variability, although we cannot decisively
rule it out \citep[see also the discussion in][]{rod11}.

As argued by \citet{mis07b}, the fact that all troughs in a mini-BAL
system vary together strongly suggests that the variability is the
result of changes in the ionization state of the absorber, presumably
because of a variable ionizing continuum.  \citet{fil14} suggest that
this mechanism may also operate in BALs.  But since the continuum
variability time scales of such luminous quasars are considerably
longer than the observed variability of mini-BALs
\citep[e.g.,][]{giv99,haw01,kas07}, one is led to invoke variable
obscuration of the ionizing continuum seen by the absorbers by
shielding material of variable column density.  This screen could be
the X-ray warm absorber \citep{gal02,gal06}, and indeed some X-ray
absorbers toward mini-BAL quasars show this particular behavior
\citep[e.g.,][]{giu11}.  In this scenario, an X-ray warm absorber lies
along sight lines toward mini-BAL and BAL systems but not toward NAL
absorbers. This conclusion is based on the measurement of high column
densities from the X-ray spectra of quasars with BALs and mini-BALs
but not from the spectra of quasars with intrinsic NALs
\citep[e.g.,][]{cha09}. Recent radiation-MHD simulations by
\citet{tak13} reproduce variable clumpy structures in warm absorbers
with typical sizes or order $20\, r_{\rm g}$ and typical optical depth
of order unity.  They appear above heights of $\sim 500\, r_{\rm g}$
from the plane of the accretion disk but, interestingly, they are not
found at very high latitudes, where sight lines towards intrinsic NAL
absorbers have been suggested to be by \citet{gan01} and
\citet{mis08}.  The fact that these clumps have sizes comparable to
the size of the continuum source, suggests that they can modulate the
apparent size of the continuum source and, by extension, the coverage
fraction of the absorbers. Thus, such clumpy structures provide a
plausible explanation for the frequent variability of mini-BALs and
the lack of frequent variability in intrinsic NALs.

\subsection{Variability of NALs and Narrow Components in mini-BAL Systems}

It is noteworthy that the intrinsic NALs we have monitored in this
program did not show any variability. These were identified as
intrinsic systems based on the fact that their \ion{C}{4}, \ion{N}{5},
or \ion{Si}{4} doublets show the signature of partial coverage of the
background source. Although the same signature is seen on some rare
occasions in low ionization lines from intervening absorbers
\citep[presumably arising in cold and compact parcels of gas;
  see][]{kob02,jon10,che13}, it has never been seen in high-ionization
lines, such as \ion{C}{4} and \ion{N}{5}, from intervening
absorbers. Therefore we are confident that the systems we have
targeted are intrinsic. Moreover, variability has been observed in
NALs in other monitoring campaigns, both at high and low redshifts, by
\citet{nar04} and \citet{wis04}. The NAL variability time scales and
resulting lower limits on electron densities from those two studies
are comparable to what we have found here for mini-BALs.

\begin{figure}
 \begin{center}
  \includegraphics[width=12cm,angle=0]{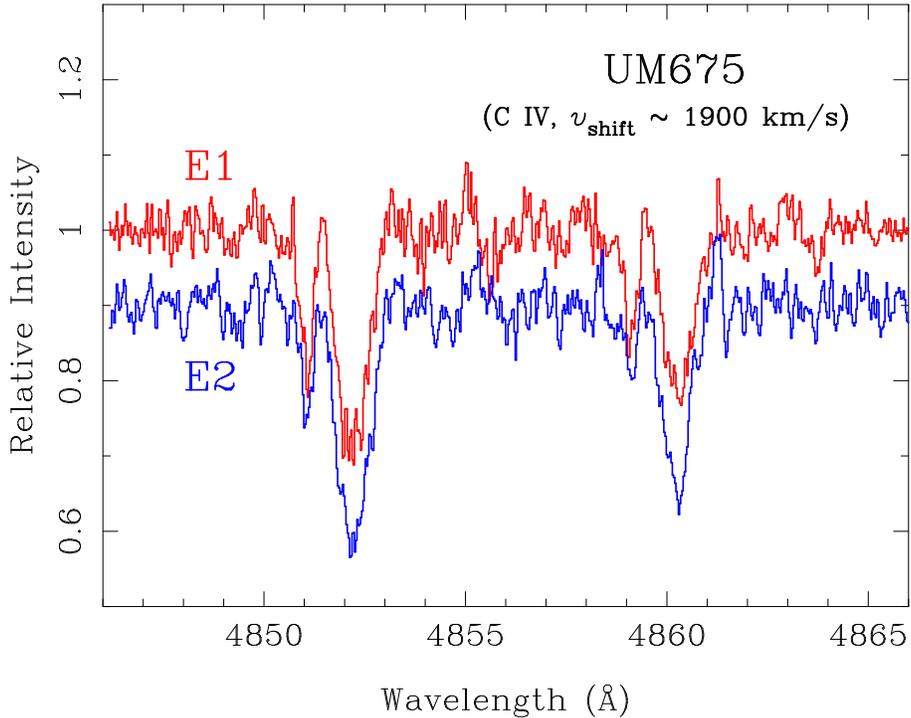}
 \end{center}
 \caption{Absorption profiles of the kinematic narrow components at
   \voff\ $\sim$ 1,900 \kms\ of the mini-BAL of UM675. The red curve
   corresponds to Epoch~1 and the blue curve to Epoch~2. The latter is
   shifted downwards by 0.1 for clarity. The broader, mini-BAL trough
   is removed by fitting it high order spline function.  Both members
   of the \ion{C}{5}$\;\lambda\lambda$1548,1551 doublet are
   shown.\label{fig:narrow}}
\end{figure}

Narrow kinematic components are sometimes also found near the centers
of mini-BALs. Examples in our sample can be seen at \voff\ $\sim
24,400\;$\kms\ in the \ion{C}{4} mini-BAL of HE1341$-$1020, at $\sim
1,300\;$\kms\ in the \ion{C}{4} and \ion{N}{5} mini-BALs of UM675, and
at $\sim 3,000\;$\kms\ in the \ion{Si}{4} mini-BAL of Q1157+014.
Among these, we can monitor the narrow components in UM675 without
complications from noise and line blending.  The narrow \ion{C}{4}
components of UM675 in Epochs 1 and 2 are shown in
Figure~\ref{fig:narrow}, in which broader absorption troughs are
fitted out by high order spline functions.  The total \ews\ (including
both \ion{C}{4}~$\lambda$1548 and \ion{C}{4}~$\lambda$1551) of Epochs
1 and 2 are measured to be $\ew({\rm E1}) = 0.58 \pm 0.02$ and
$\ew({\rm E2}) = 0.56 \pm0.01$. Thus, they show no variability over a
time interval of $\deltrest \sim 3.5$~ years. \citet{ham97b} noted a
similar behavior for the narrow core of the \ion{C}{4} mini-BAL in
Q2343+125 and suggested the narrow components may not be physically
associated to the quasar itself, but arising in the quasar host galaxy
or foreground galaxy (although this quasar is one of our targets, we
cannot monitor the variability of the narrow components because our
spectra have lower \sn\ than those of \citet{ham97b}).  However,
narrow components appear to always coincide with the centers of
mini-BAL profiles, which suggests a physical association between the
absorbing media. One possible explanation for the behavior of the
narrow cores of mini-BALs is that they originate in dense clumps of
high optical depth, embedded in the mini-BAL flow. When the ionizing
continuum varies, the broad mini-BAL troughs show an easily
discernible response but the highly saturated, narrow cores do not.

\subsection{Possible Geometry and Locations of mini-BAL and NAL Gas
\label{sec:geom}}

The variability properties of mini-BALs and NALs reported here in
combination with previous reports in the literature (see references
above), lead us to suggest the following possibilities for locations
of the absorbing gas. These are inspired by numerical simulations of
quasar outflows (see references in \S\ref{sec:intro}) and are a
refinement of the picture suggested by \citet{gan01}.

\begin{figure}
 \begin{center}
  \includegraphics[width=15cm,angle=0]{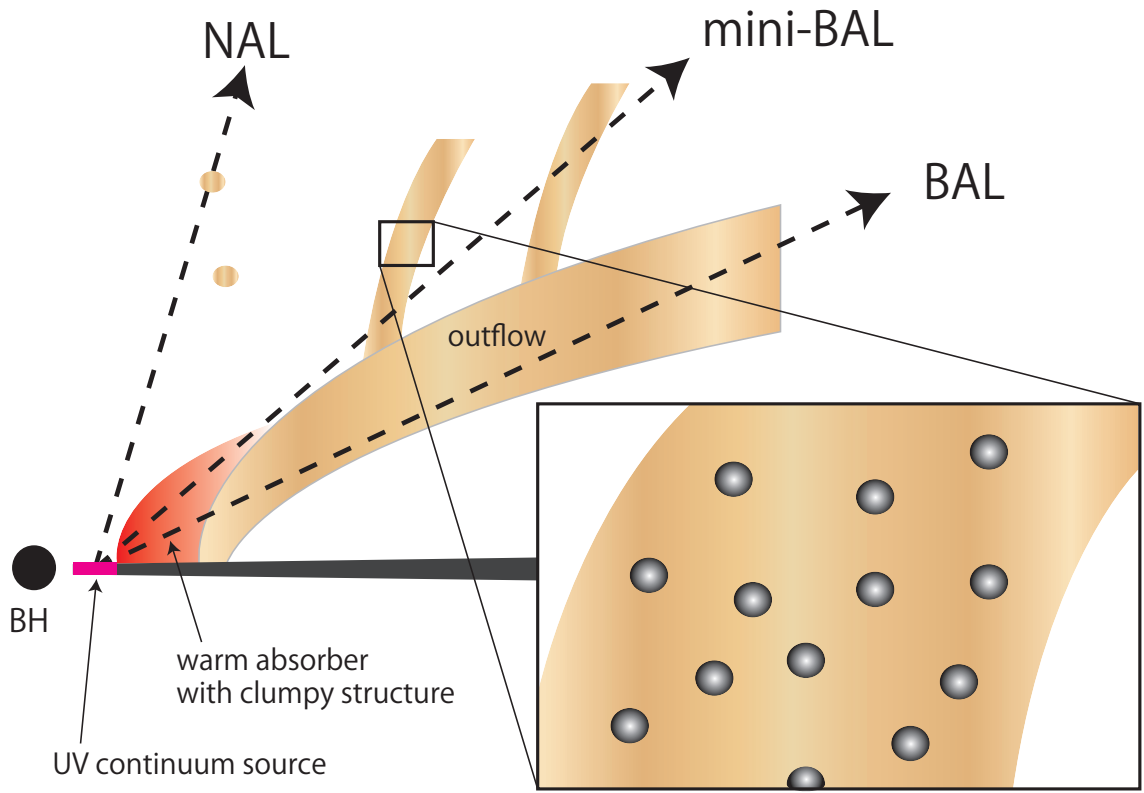}
 \end{center}
 \caption{Possible geometry of the locations of NAL, mini-BAL, and BAL
   absorbers in the accretion-disk wind scenario (see text in
   detail).\label{fig:geometry}}
\end{figure}

In the first scenario, both types of absorbers are located in an
accretion-disk wind, as illustrated in Figure~\ref{fig:geometry}.  The
mini-BAL gas is in the form of filaments rising from the fast
component of the outflow. They make up the interface between the dense
BAL wind at low latitudes above the disk and a hotter, much more
tenuous medium in the polar direction. As such, the mini-BAL filaments
have a large outflow speed, comparable to that of the BAL gas and a
substantial velocity gradient. Within those filaments there may exist
small, dense clumps that are sometimes observed as narrow cores in
mini-BAL troughs. Variability of the ionizing continuum incident on
the mini-BAL filaments can change the optical depth of the mini-BAL
troughs arising in the filaments but not those of their narrow cores
which arise in the dense clumps with \ion{C}{4} optical depths of
$\tau_{\rm C\;IV}\gg 1$ (assuming their predicted hydrogen column
densities of $\gtrsim 10^{20}\;{\rm cm}^{-2}$, $\log U\sim -1.5$, and
solar abundances). In this scenario, the NALs and the narrow cores of
mini-BALs arise in higher-density, smaller filaments in the same
region of the wind or at higher latitudes. It is possible that the NAL
filaments are denser than the mini-BAL filaments thus the NALs are
less susceptible to fluctuations in the ionizing continuum.  However,
NALs could vary because of changes in the coverage fraction, which
could come about as the apparent size of the background source changes
because of fluctuations in the structure of the inner screen (see
discussion at the end of \S\ref{sec:geom}).

Alternatively, the NAL gas may be located in the host galaxy, at large
distances from the center, as we noted in \S\ref{sec:intro}. In this
scenario, small NAL filaments may form from the interaction of the
wind's blast wave with cold clouds in the host galaxy, as proposed by
\citet{fau12}. Radiative shocks resulting from this interaction can
create thin, dense filaments whose properties are consistent with what
is inferred from photoionization models for the NAL absorbers
\citep[e.g.,][]{wu10,borg13} and are likely to have $\tau_{\rm
  C\;IV}\gg 1$. Thus the variability of the NALs arising in these
filaments is more likely to result from fluctuations in the coverage
fraction than fluctuations in the ionizing continuum. This scenario is
more likely to apply to {\it associated} NALs, with $\voff \le
5,000\;\kms$, as this range of blueshifts better matches the
prediction of numerical models for the interaction of quasar outflows
with the ISM of the host galaxy \citep{kuro09}. NALs found at $\voff
\gg 5,000\;\kms$ \citep[e.g., 20,000--65,000~\kms; see][]{mis07a} are
better explained by the previous scenario.

The underlying hypotheses in the above scenarios are that (a) NAL
variations are more rare than those of mini-BALs (although when NALs
do vary they do so on the same time scales as mini-BALs), and (b) that
NAL variability is caused primarily by fluctuations in the coverage
fraction while the variability of mini-BALs can be caused by
fluctuations in both the coverage fraction and the ionizing continuum.
Both of these hypotheses warrant further testing, which can be done by
a more detailed analysis of the data we have presented in this paper
as well as data collected from denser monitoring campaigns on select
targets.

\section{SUMMARY}

We have monitored 12 intrinsic NALs (selected by their partial
coverage signature) and 7 mini-BALs in quasars of $z_{em}\approx 2$--3
for $\sim4$--12 years ($\sim1$--3.5 years in the quasar
rest-frame). Using high-dispersion spectra, we have measured the line
\ews\ as a function of time and have looked for substantial variations
in the line profiles. Our main results are as follows.

\begin{itemize}
\item We detect significant variability in the \ews\ of mini-BALs
  without any discernible changes in their profiles. However, we do
  not detect any significant variability in the NALs nor in the narrow
  cores that are sometimes present in mini-BAL profiles. 
\item The \ew\ variability of mini-BALs is larger on longer time
  scales, a behavior that has also been seen in BALs.
\item When we are able to measure variations in the \ew\ of more than
  one transition at the same time (\ion{C}{4}, \ion{Si}{4},
  \ion{N}{5}) the changes occur in the same direction.  This pattern
  has also been observed in BALs and is consistent with a scenario in
  which the variations result from changes in the ionizing continuum.
\item If we assume that the variability timescale is the recombination
  time, we can place lower limits on the electron density of the
  mini-BAL gas of $\sim 10^3$--$10^4\;$\cmm. These imply upper limits
  on the distances of the absorbers in the range 3--6~kpc, which could be 
  reduced by a factor of $\sim 4$ for a different choice of quasar SED.
\item The observational results lead us to suggest two possible
  scenarios for the structure and location of the mini-BAL and NAL
  gas. In the first scenario both types of absorber are associated
  with the high-latitude potion of an accretion-disk wind while in
  the second, the NAL gas is at distances $\gtrsim 100\;$pc from the
  quasar central engine.
\end{itemize}

The next stage of this work will entail a more detailed analysis of
the variability data to study the behavior of the coverage fractions
and the relative strengths of different lines, and interpret these
with the help of photoionization models. Further progress in
characterizing and interpreting the variability properties of
mini-BALs and NALs can be made via dense monitoring campaigns of
select targets in which we can observe changes on shorter time scales
than we were able to probe here and associate them with changes in the
ionizing continuum (via X-ray and UV observations). The data from such
campaigns can also help us place better constraints on the scenario in
which variability is cause by parcels of gas crossing the cylinder of
sight to the continuum source.

It would also be extremely useful to quantify the properties that the
variable shield we have invoked must have in order to produce the
desired behavior. There are a number of constraints that such a shield
must satisfy: it must reproduce the low column densities observed in
the X-ray spectra, the continuum it transmits must lead to the
observed ionization state of the UV absorber, and its structure must
allow for variations in far-UV transparency and apparent source size
to vary in order to reproduce the observed mini-BAL variability. It is
also worth considering further different geometric arrangements of the
shield, the absorber, and the background source(s) relative to the
cylinder of sight, specifically whether the source of ionizing
photons, the shield, and the mini-BAL gas must all be aligned.

\acknowledgments We would like to thank David Tytler and Fred Hamann
for providing us with their Keck/HIRES data.  We also thank the
anonymous referee for a number of comments that helped us improve the
paper. The research was supported by JGC-S Scholarship Foundation.
This work was also supported by NASA grant NAG5-10817 and by NSF
grants AST-0807993 and AST-1312686 at Penn State. M.E. would like to
thank the Center for Relativistic Astrophysics at Georgia Tech and the
Department of Astronomy at the University of Washington for their warm
hospitality during the writing of this manuscript.

\clearpage


\begin{deluxetable}{crrllcrccc}
\tabletypesize{\scriptsize}
\setcounter{table}{0}
\tablecaption{List of Monitored Quasars \label{tab:quasars}}
\tablewidth{0pt}
\tablehead{
\colhead{QSO}       &
\colhead{RA$^a$}     &
\colhead{Dec$^a$}    &
\colhead{$m_{V}^b$}  &
\colhead{$z_{em}^c$} &
\colhead{$z_{abs}^c$} &
\colhead{\voff$^d$}   &
\colhead{Class$^e$}   &
\colhead{variability$^f$}  &
\colhead{reference$^g$}  \\
\colhead{}          &
\colhead{[hh:mm:ss]}&
\colhead{[dd:mm:ss]}&
\colhead{[mag.]}    &
\colhead{}          &
\colhead{}          &
\colhead{[\kms]}    &
\colhead{}          &
\colhead{}          &
\colhead{}          \\
\colhead{(1)}       &
\colhead{(2)}       &
\colhead{(3)}       &
\colhead{(4)}       &
\colhead{(5)}       &
\colhead{(6)}       &
\colhead{(7)}       &
\colhead{(8)}       &
\colhead{(9)}       &
\colhead{(10)}
}
\startdata
\hline
\multicolumn{9}{c}{NAL} \\
\hline
HE0130$-$4021 & 01:33:01.9 & $-$40:06:28 & 17.02 & 3.030    &       2.5597 &  37037      & A & N & 1 \\  
              &            &             &       &          &       2.9749 &   4129      & A & N & 1 \\  
              &            &             &       &          &       2.2316 &  65181      & B & N & 1 \\  
HE0940$-$1050 & 09:42:53.4 & $-$11:04:25 & 16.90 & 3.080    &       2.8347 &  18578      & B & N & 1 \\  
Q1009+2956    & 10:11:55.6 & $+$29:41:42 & 16.05 & 2.644    &       2.2533 &  33879      & A & N & 1 \\  
              &            &             &       &          &       2.6495 &    452      & A & N & 1 \\  
HS1700+6416   & 17:01:00.6 & $+$64:12:09 & 16.17 & 2.722    &       2.7125 &    767      & A & N & 1 \\  
              &            &             &       &          &       2.7164 &    452      & A & N & 1 \\  
              &            &             &       &          &       2.4330 &  24195      & B & N & 1 \\  
              &            &             &       &          &       2.4394 &  23640      & B & N & 1 \\  
HS1946+7658   & 19:44:55.0 & $+$77:05:52 & 16.20 & 3.051    &       3.0385 &    927      & A & N & 1 \\  
              &            &             &       &          &       3.0497 &     96      & A & N & 1 \\
\hline
\multicolumn{9}{c}{mini-BAL} \\
\hline
UM675         & 01:52:27.3 & $-$20:01:07 & 17.40 & 2.15     & $\sim$2.13   &  $\sim$1,900 & A & Y & 2 \\
HE0151$-$4326 & 01:53:27.2 & $-$43:11:38 & 16.80 & 2.78$^h$ & $\sim$2.64   &  $\sim$11,300& A & Y & 3 \\
Q1157+014     & 11:59:44.8 & $+$01:12:07 & 17.52 & 1.9997   & $\sim$1.97   &  $\sim$3,000 & A & Y & 4 \\
HE1341$-$1020 & 13:44:27.1 & $-$10:35:42 & 17.80 & 2.135    & $\sim$2.12   &  $\sim$1,300 & A & Y & 3 \\
              &            &             &       &          & $\sim$2.67   &  $\sim$8,900 & A & Y & 3 \\
HS1603$+$3820 & 16:04:55.4 & $+$38:12:02 & 15.99 & 2.542$^i$& $\sim$2.43   &  $\sim$9,500 & A & Y$^i$ & 5 \\
Q2343+125     & 23:46:28.2 & $+$12:49:00 & 17.00 & 2.515    & $\sim$2.24   & $\sim$24,400 & A & N & 6 \\
\enddata 
\tablenotetext{a}{Coordinates (J2000) from NASA/IPAC Extragalactic
  Database (NED).}
\tablenotetext{b}{V magnitude from \citet{ver10}.}
\tablenotetext{c}{Quasar emission redshift and absorption redshift of
  NAL/mini-BAL system that we monitored. These are from the reference
  listed in (10).}
\tablenotetext{d}{Ejection velocity from the quasar emission redshift,
  calculated via
  $\voff=c\,[(1+z_{em})^2-(1+z_{abs})^2]\,/\,[(1+z_{em})^2+(1+z_{abs})^2]$,
  where $z_{em}$ and $z_{abs}$ are the emission redshift of the quasar
  and the absorption redshift of the NAL/mini-BAL, respectively. A
  positive value indicates a blueshifted line.}
\tablenotetext{e}{Reliability class: A and B mean reliable and
  possibly intrinsic absorption lines, respectively
  \citep[see][]{mis07a}.}
\tablenotetext{f}{NAL/mini-BAL varies or not.}
\tablenotetext{g}{References: (1) \citet{mis07a}, (2) \citet{ham95},
  (3) \citet{nar07}, (4) \citet{wri79}, (5) \citet{dob99}, (6)
  \citet{ham97b},.}
\tablenotetext{h}{The emission redshift is listed as $z_{em}=2.74$ in
  \citet{nar07}, but we adopt $\sim$2.78 because the Ly$\alpha$ forest
  starts to appear at $\lambda$ $\sim$4600\AA.}
\tablenotetext{i}{From \citet{mis07b}.}
\end{deluxetable}
\clearpage

\begin{deluxetable}{ccclccrcc}
\tabletypesize{\scriptsize}
\tablecaption{Log of Monitoring Observations \label{tab:obs}}
\tablewidth{0pt}
\tablehead{
\colhead{QSO}                &
\colhead{Type}               &
\colhead{Epoch}              &
\colhead{Obs. Date}          &
\colhead{Instrument}         &
\colhead{$\lambda$-coverage$^a$} &
\colhead{Exp. Time}          &
\colhead{$R$}                &
\colhead{Note$^b$}           \\
\colhead{}                   &
\colhead{}                   &
\colhead{}                   &
\colhead{}                   &
\colhead{}                   &
\colhead{[\AA]}              &
\colhead{[sec]    }          &
\colhead{}                   &
\colhead{}                   \\
\colhead{(1)}                &
\colhead{(2)}                &
\colhead{(3)}                &
\colhead{(4)}                &
\colhead{(5)}                &
\colhead{(6)}                &
\colhead{(7)}                &
\colhead{(8)}                &
\colhead{(9)} 
}
\startdata
HE0130$-$4021 &  NAL     &  1  & 1995 Dec 28        &  Keck+HIRES  &  3700     -- 6062     &  14400  &       36,000     &   $^c$         \\
              &          &  2  & 2003 Jan 13        &  VLT+UVES    &  3523     -- 4519     &   3065  & $\sim$40,000$^d$ &   70.B-0522(A) \\
              &          &     &                    &              &  4782     -- 5751     &         &                  &                \\
              &          &     &                    &              &  5839     -- 6264     &         &                  &                \\
              &          &  3  & 2007 Sep 5         &  VLT+UVES    &  3508     -- 4507     &   6000  & $\sim$40,000$^d$ &  079.B-0469(A) \\
              &          &     &                    &              &  4629     -- 5593     &         &                  &                \\
              &          &     &                    &              &  5680     -- 6349$^e$ &         &                  &                \\
\hline
HE0940$-$1050 &  NAL     &  1  & 2000 Apr 03        &  VLT+UVES    &  3601     -- 3870     &   3600  & $\sim$40,000$^d$ &   65.O-0474(A) \\
              &          &     &                    &              &  4783     -- 5756     &         &                  &                \\
              &          &     &                    &              &  5837     -- 6428$^e$ &         &                  &                \\
              &          &  2  & 2001 Feb 02--14    &  VLT+UVES    &  3605     -- 5750     &  28800  & $\sim$40,000$^d$ &  166.A-0106(A) \\
              &          &     &                    &              &  5847     -- 6428$^e$ &         &                  &                \\
              &          &  3  & 2007 Jun 6,7       &  VLT+UVES    &  3588     -- 4505     &   6000  & $\sim$40,000$^d$ &  079.B-0469(A) \\
              &          &     &                    &              &  4630     -- 5591     &         &                  &                \\
              &          &     &                    &              &  5683     -- 6428$^e$ &         &                  &                \\
\hline
Q1009+2956    &  NAL     &  1  & 1995 Dec 29        &  Keck+HIRES  &  4303     -- 5517     &  12200  &       36,000     &   $^c$         \\
              &          &  2  & 1998 Dec 15        &  Keck+HIRES  &  4350     -- 4612     &  14260  &       36,000     &   $^c$         \\
              &          &  3  & 1999 Mar 09        &  Keck+HIRES  &  4300     -- 4642     &  14200  &       36,000     &   $^c$         \\
              &          &  4  & 2007 Jun 5         &  VLT+UVES    &  4169     -- 5158     &   3000  & $\sim$40,000$^d$ &  079.B-0469(A) \\
              &          &     &                    &              &  5238     -- 5741$^e$ &         &                  &                \\
\hline
HS1700+6416   &  NAL     &  1  & 1994 Apr 07        &  Keck+HIRES  &  4490     -- 5850$^e$ &   6750  &       36,000     &   $^c$         \\
              &          &  2  & 1995 May 10        &  Keck+HIRES  &  4460     -- 5864$^e$ &  11500  &       36,000     &   $^c$         \\
              &          &  3  & 2005 Jul 07        &  Subaru+HDS  &  3547     -- 4900     &   3600  &       36,000     &   S05A-041     \\
              &          &     &                    &              &  4991     -- 5864$^e$ &         &                  &                \\
              &          &  4  & 2005 Aug 19        &  Subaru+HDS  &  3547     -- 4897     &   5400  &       36,000     &   S05A-041     \\
              &          &     &                    &              &  4992     -- 5864$^e$ &         &                  &                \\
\hline
HS1946+7658   &  NAL     &  1  & 1994 Jul 31        &  Keck+HIRES  &  4860     -- 6296$^e$ &  25200  &       36,000     &   $^c$         \\
              &          &  2  & 1998 Sep 25        &  Keck+HIRES  &  4800     -- 5047     &   8000  &       36,000     &   $^c$         \\
              &          &  3  & 1998 Oct 26--28    &  Keck+HIRES  &  4850     -- 5020     &  24000  &       36,000     &   $^c$         \\
\hline
UM675         & mini-BAL &  1  & 1994 Sep 24,25     &  Keck+HIRES  &  3750     -- 4963$^e$ &  18000  &       34,000     &   $^f$         \\
              &          &  2  & 2005 Aug 19,20     &  Subaru+HDS  &  3600     -- 4963$^e$ &  23000  &       36,000     &   S05A-041     \\
\hline
HE0151$-$4326 & mini-BAL &  1  & 2001 Sep 18--22    &  VLT+UVES    &  3160     -- 5752     &  18000  & $\sim$40,000$^d$ &  166.A-0106(A) \\
              &          &     &                    &              &  5836     -- 5893$^e$ &         &                  &                \\
              &          &  2  & 2001 Oct 09,10     &  VLT+UVES    &  3161     -- 3869     &  14400  & $\sim$40,000$^d$ &  166.A-0106(A) \\
              &          &     &                    &              &  4785     -- 5755     &         &                  &                \\
              &          &     &                    &              &  5842     -- 5893$^e$ &         &                  &                \\
              &          &  3  & 2001 Nov 16--19    &  VLT+UVES    &  3161     -- 5752     &  28800  & $\sim$40,000$^d$ &  166.A-0106(A) \\
              &          &     &                    &              &  5846     -- 5893$^e$ &         &                  &                \\
              &          &  4  & 2007 Sep 5         &  VLT+UVES    &  3486     -- 4504     &   3000  & $\sim$40,000$^d$ &  079.B-0469(A) \\
              &          &     &                    &              &  4789     -- 5753     &         &                  &                \\
              &          &     &                    &              &  5855     -- 5893$^e$ &         &                  &                \\
\hline
Q1157+014     & mini-BAL &  1  & 2000 Apr 30        &  VLT+UVES    &  3640     -- 3870     &   7200  & $\sim$40,000$^d$ &   65.O-0063(B) \\
              &          &     &                    &              &  4542     -- 4727$^e$ &         &                  &                \\
              &          &  2  & 2001 Jun 14,15     &  VLT+UVES    &  3636     -- 4411     &  10800  & $\sim$40,000$^d$ &   67.A-0078(A) \\
              &          &  3  & 2002 Jan 23        &  VLT+UVES    &  3715     -- 4507     &   1800  & $\sim$40,000$^d$ &   68.A-0461(A) \\
              &          &  4  & 2006 Jan 23        &  VLT+UVES    &  3643     -- 4727$^e$ &   1960  & $\sim$40,000$^d$ &  076.A-0860(A) \\
              &          &  5  & 2007 Jun 6,7       &  VLT+UVES    &  3762     -- 4727$^e$ &   6000  & $\sim$40,000$^d$ &  079.B-0469(A) \\
\hline
HE1341$-$1020 & mini-BAL &  1  & 2001 May 02--12    &  VLT+UVES    &  3251     -- 4938$^e$ &  43200  & $\sim$40,000$^d$ &  166.A-0106(A) \\
              &          &  2  & 2007 Aug 4,5       &  VLT+UVES    &  3490     -- 4510     &   6000  & $\sim$40,000$^d$ &  079.A-0469(A) \\
              &          &     &                    &              &  4629     -- 4938$^e$ &         &                  &                \\
\hline
HS1603+3820   & mini-BAL &  1  & 2002 Mar 23        &  Subaru+HDS  &  5080     -- 5581$^e$ &   2700  & $\sim$45,000     &   S01B-057     \\
              &          &  2  & 2003 Jul 7         &  Subaru+HDS  &  3520     -- 4850     &   6000  & $\sim$45,000     &   Eng. Time.   \\
              &          &     &                    &              &  4930     -- 5581$^e$ &         &                  &                \\
              &          &  3  & 2005 Feb 26        &  Subaru+HDS  &  3520     -- 4855     &   7100  & $\sim$36,000     &   S04B-003     \\
              &          &     &                    &              &  4925     -- 5581$^e$ &         &                  &                \\
              &          &  4  & 2005 Jun 29        &  Subaru+HDS  &  3520     -- 4855     &   3600  & $\sim$45,000     &   S05A-041     \\
              &          &     &                    &              &  4925     -- 5581$^e$ &         &                  &                \\
              &          &  5  & 2005 Aug 19        &  Subaru+HDS  &  3520     -- 4850     &   3600  & $\sim$36,000     &   S05A-041     \\
              &          &     &                    &              &  4920     -- 5581$^e$ &         &                  &                \\
              &          &  6  & 2006 May 31--Jun 1 &  Subaru+HDS  &  4320     -- 5581$^e$ &   9000  & $\sim$45,000     &   S06A-153S    \\
\hline
Q2343+125     & mini-BAL &  1  & 2002 Aug 18        &  VLT+UVES    &  3409     -- 3872     &   7200  & $\sim$40,000$^d$ &   69.A-0204(A) \\
              &          &     &                    &              &  4785     -- 5538$^e$ &         &                  &                \\
              &          &  2  & 2003 Oct 29,30     &  VLT+UVES    &  3526     -- 4507     &   7200  & $\sim$40,000$^d$ &  072.A-0346(A) \\
              &          &     &                    &              &  4801     -- 5538$^e$ &         &                  &                \\
              &          &  3  & 2007 Sep 16        &  VLT+UVES    &  4192     -- 5157     &   3000  & $\sim$40,000$^d$ &  079.B-0469(A) \\
              &          &     &                    &              &  5240     -- 5538$^e$ &         &                  &                \\
\enddata
\tablenotetext{a}{Wavelength range covered by the observed
  spectrum. Small gaps of $\leq$20~\AA\ that are sometimes seen in
  Keck/HIRES spectra are ignored.}
\tablenotetext{b}{Proposal ID of the relevant observing program.}
\tablenotetext{c}{Provided by David Tytler.}
\tablenotetext{d}{Resolving power-slit products (i.e., $R \times
  \Delta\theta$ [arcsec]) are 41,400 at 4500\AA\ on Blue Arm and
  38,700 at 6000\AA\ on Red Arm. If we use the 1\farcs0 slit, the
  resolving power is $\sim$40,000.}
\tablenotetext{e}{Maximum effective wavelength, which corresponds to
  5000~\kms\ redward from \ion{C}{4} emission line. At higher
  wavelength than this limit, no \ion{C}{4}, \ion{Si}{4}, and
  \ion{N}{5} NALs/mini-BALs are expected to exist.}
\tablenotetext{f}{Provided by Fred Hamann.}
\end{deluxetable}
\clearpage

\begin{deluxetable}{ccccccc}
\tabletypesize{\scriptsize}
\tablecaption{Observed Frame Equivalent Width \label{tab:ew}}
\tablewidth{0pt}
\tablehead{
\multicolumn{3}{c}{}                   &
\colhead{}                             &
\multicolumn{3}{c}{$\ew_{\rm obs}$}       \\
\cline{5-7}
\colhead{QSO}                          &
\colhead{\voff}                        &
\colhead{Epoch}                        &
\colhead{$\Delta t^a$}                 &
\colhead{\ion{N}{5}}                   &
\colhead{\ion{Si}{4}}                  &
\colhead{\ion{C}{4}}                   \\
\colhead{}                             &
\colhead{[\kms]}                       &
\colhead{}                             &
\colhead{[days]}                        &
\colhead{[\AA]}                        &
\colhead{[\AA]}                        &
\colhead{[\AA]}                        \\
\colhead{(1)}                          &
\colhead{(2)}                          &
\colhead{(3)}                          &
\colhead{(4)}                          &
\colhead{(5)}                          &
\colhead{(6)}                          &
\colhead{(7)}                          
}
\startdata
HE0130$-$4021 &      65,181 &  1  &    0 & $^b$                     & $^b$                  & 0.99$\pm$0.05           \\
              &            &  2  & 2573 & $^b$                     & $^b$                  & 0.99$\pm$0.07           \\
              &            &  3  & 4269 & $^b$                     & $^b$                  & 1.017$\pm$0.04          \\
              \cline{2-7}                
              &      37,037 &  1  &    0 & $^b$                     & 0.057$\pm$0.008$^f$   & 1.49$\pm$0.08           \\
              &            &  2  & 2573 & $^b$                     & 0.059$\pm$0.013       & 1.46$\pm$0.12           \\
              &            &  3  & 4269 & $^b$                     & 0.054$\pm$0.007       & 1.45$\pm$0.07           \\
              \cline{2-7}                
              &       4,129 &  1  &    0 & 1.87$\pm$0.14            &  $<$0.23              & -                       \\
              &            &  2  & 2573 & 1.82$\pm$0.18            &  $<$0.28              & 1.17$\pm$0.25$^f$       \\
              &            &  3  & 4269 & 1.73$\pm$0.11$^f$        &  $<$0.17              & 1.15$\pm$0.15           \\
\hline                                                             
HE0940$-$1050 &      18,578 &  1  &    0 &   -                      & 0.538$\pm$0.028        & 1.62$\pm$0.06          \\
              &            &  2  &  311 &   $^b$                   & 0.533$\pm$0.014        & 1.65$\pm$0.03          \\
              &            &  3  & 2620 &   $^b$                   & 0.542$\pm$0.021        & 1.61$\pm$0.05          \\
\hline                                                             
Q1009+2956    &      33,879 &  1  &    0 &   -                      & 0.900$\pm$0.030        & 1.73$\pm$0.03          \\
              &            &  2  & 1082 &   -                      & 0.908$\pm$0.003        &   -                    \\
              &            &  3  & 1166 &   -                      & 0.895$\pm$0.014        &   -                    \\
              &            &  4  & 4176 &   -                      & 0.926$\pm$0.048        & 1.72$\pm$0.06          \\
              \cline{2-7}                                          
              &        452 &  1  &    0 & 0.434$\pm$0.008$^f$      & 0.019$\pm$0.014        &   -                    \\
              &            &  2  & 1082 & 0.392$\pm$0.005          &   -                    &   -                    \\
              &            &  3  & 1166 & 0.405$\pm$0.004          &   -                    &   -                    \\
              &            &  4  & 4176 & 0.399$\pm$0.016          & 0.018$\pm$0.009        & 0.430$\pm$0.015        \\
\hline                                                             
HS1700+6416   &      24,195 &  1  &    0 &   -                      & 0.102$\pm$0.019        & 0.406$\pm$0.033        \\
              &            &  2  &  398 &   -                      & 0.104$\pm$0.006        & 0.424$\pm$0.014        \\
              &            &  3  & 4109 &  $^b$                    & 0.084$\pm$0.015        & 0.430$\pm$0.036$^f$    \\
              &            &  4  & 4152 &  $^b$                    & 0.106$\pm$0.01         & 0.417$\pm$0.027        \\
              \cline{2-7}                                          
              &      23,640 &  1  &    0 &   -                      & 0.087$\pm$0.017       & 0.196$\pm$0.019$^f$     \\
              &            &  2  &  398 &   -                      & 0.100$\pm$0.007       & 0.216$\pm$0.007$^f$     \\
              &            &  3  & 4109 &  $^b$                    & 0.085$\pm$0.015       & 0.189$\pm$0.018$^f$     \\
              &            &  4  & 4152 &  $^b$                    & 0.089$\pm$0.014       & 0.197$\pm$0.012         \\
              \cline{2-7}                                          
              &        767 &  1  &    0 & 0.501$\pm$0.027          &  $<$0.02              & 0.291$\pm$0.016         \\
              &            &  2  &  398 & 0.465$\pm$0.009          &  $<$0.007             & 0.295$\pm$0.007         \\
              &            &  3  & 4109 & 0.500$\pm$0.031          &  $<$0.02              & 0.328$\pm$0.021         \\
              &            &  4  & 4152 & 0.504$\pm$0.022          &  $<$0.01              & 0.318$\pm$0.015         \\
              \cline{2-7}                                          
              &        452 &  1  &    0 & 0.108$\pm$0.015          &  $<$0.015             &  $<$0.011               \\
              &            &  2  &  398 & 0.101$\pm$0.006          &  $<$0.006             &  $<$0.005               \\
              &            &  3  & 4109 & 0.084$\pm$0.015          &  $<$0.02              &  $<$0.02                \\
              &            &  4  & 4152 & 0.109$\pm$0.012          &  $<$0.01              &  $<$0.01                \\
\hline                                                             
HS1946+7658   &        927 &  1  &    0 &  $<$0.01                 &  $<$0.01              & 0.286$\pm$0.007         \\
              &            &  2  & 1517 &  $<$0.01                 &   -                   &   -                     \\
              &            &  3  & 1549 &  $<$0.01                 &   -                   &   -                     \\
              \cline{2-7}                                          
              &         96 &  1  &    0 & 0.366$\pm$0.008          & 2.318$\pm$0.01        & 4.164$\pm$0.02          \\
              &            &  2  & 1517 & 0.384$\pm$0.014          &   -                   &   -                     \\
              &            &  3  & 1549 & 0.371$\pm$0.012          &   -                   &   -                     \\
\hline                                                             
UM675         & $\sim$1,900 &  1  &    0 & 6.15$\pm$1.01            &  $<$0.4               & 5.38$\pm$0.24           \\
              &            &  2  & 3982 & 3.63$\pm$0.33            &  $<$0.3               & 3.84$\pm$0.21           \\
\hline                                                             
HE0151$-$4326 & $\sim$11,300&  1  &    0 &   $^b$                   &  $<$0.07              & 1.96$\pm$0.26           \\
              &            &  2  &   19 &   -                      &  $<$0.10              & 2.12$\pm$0.37           \\
              &            &  3  &   58 &   $^b$                   &  $<$0.10              & 2.22$\pm$0.36           \\
              &            &  4  & 2176 &   $^b$                   & 0.24$\pm$0.16         & 3.69$\pm$0.53           \\
              \cline{2-7}                                          
              & $\sim$8,900 &  1  &    0 &   $^b$                   & 0.09$\pm$0.04         & 4.07$\pm$0.21           \\
              &            &  2  &   19 &   -                      & 0.10$\pm$0.06         & 4.22$\pm$0.30           \\
              &            &  3  &   58 &   $^b$                   & 0.08$\pm$0.06         & 3.92$\pm$0.32           \\
              &            &  4  & 2176 &   -                      & 0.59$\pm$0.09         & 10.42$\pm$0.43          \\
\hline                                                             
Q1157+014     & $\sim$3,000 &  1  &    0 & 38.38$\pm$1.28           &   -                   & 37.12$\pm$1.23          \\
              &            &  2  &  410 & 38.27$\pm$1.17           &  9.16$\pm$0.52$^c$    &   -                     \\
              &            &  3  &  633 &   -                      & 10.31$\pm$1.23$^c$    &   -                     \\
              &            &  4  & 2094 & 37.7$\pm$4.5             &  9.77$\pm$1.12$^c$    & 38.17$\pm$2.33          \\
              &            &  5  & 2593 &   -                      & 11.26$\pm$0.71$^c$    & 38.53$\pm$1.05          \\
\hline                                                             
HE1341$-$1020 & $\sim$1,300 &  1  &    0 & 15.39$\pm$0.47           &  $<$0.3               & 6.86$\pm$0.24           \\
              &            &  2  & 2280 & 21.63$\pm$0.79           &  $<$0.7               & 12.40$\pm$0.45          \\
\hline                                                             
HS1603+3820   & $\sim$9,500 &  1  &    0 &   -                      &   -                   & 10.12$\pm$1.15          \\
              &            &  2  &  471 &   $^b$                   &  $^d$                 & 17.95$\pm$1.84          \\
              &            &  3  & 1071 &   $^b$                   &  $<$0.8$^e$           & 13.62$\pm$0.79          \\
              &            &  4  & 1194 &   $^b$                   &  $^d$                 & 13.03$\pm$1.43          \\
              &            &  5  & 1245 &   $^b$                   &  $^d$                 & 12.69$\pm$0.84          \\
              &            &  6  & 1530 &   -                      &  $<$0.5               & 13.34$\pm$0.61          \\
\hline                                                             
Q2343+125     & $\sim$24,400&  1  &    0 &   -                      &   -                   & 3.29$\pm$0.69           \\
              &            &  2  &  437 &   $^b$                   &   -                   & 2.04$\pm$0.44           \\
              &            &  3  & 1855 &   -                      &  $<$0.76              & 2.69$\pm$0.63           \\
\enddata
\tablenotetext{a}{Time delay from the first epoch in the observed frame.}
\tablenotetext{b}{This line is blended with the Ly$\alpha$ forest.}
\tablenotetext{c}{Parameters are measured using only blue component, because doublet lines are not self-blended and can be fitted separately.}
\tablenotetext{d}{This line is significantly affected by bad columns.}
\tablenotetext{e}{Spectral region is partially truncated.}
\tablenotetext{f}{This line is affected by a data defect that could slightly alter the \ew\ measurement.}
\tablenotetext{-}{This line is not covered by the observed spectrum.}
\end{deluxetable}
\clearpage

\begin{deluxetable}{ccccccccc}
\tabletypesize{\scriptsize}
\tablecaption{Variability and Physical Properties of mini-BAL Absorbers \label{tab:prop}}
\tablewidth{0pt}
\tablehead{
\colhead{QSO}       &
\colhead{\voff}     &
\colhead{${t_{var}}^a$}&
\colhead{${S_\ew}^b$}&
\colhead{ion}&
\colhead{Ionization}&
\colhead{${L_{\rm bol}}^c$}   &
\colhead{${n_e}^d$} &
\colhead{$r^e$ }   \\
\colhead{}               &
\colhead{[\kms]}         &
\colhead{[days]}         &
\colhead{}               &
\colhead{}               &
\colhead{Change}         &
\colhead{[erg~s$^{-1}$]} &
\colhead{[cm$^{-3}$]}     &
\colhead{[kpc]}           \\
\colhead{(1)}       &
\colhead{(2)}       &
\colhead{(3)}       &
\colhead{(4)}       &
\colhead{(5)}       &
\colhead{(6)}       &
\colhead{(7)}       &
\colhead{(8)}       &
\colhead{(9)}       
}
\startdata
UM675         & $\sim$1900 & 3982 (E1 $\rightarrow$ E2) & $-2.4 $    & \ion{N}{5}  & \ion{N}{5}  $\rightarrow$ \ion{N}{4}   & $3.8\ten{47}$ & $\geq 1.66\ten{3}$ & $\leq 5.6$ \\
              &            &                            & $-4.8 $    & \ion{C}{4}  & \ion{C}{4}  $\rightarrow$ \ion{C}{3}   &               & $\geq 3.27\ten{3}$ & $\leq 4.0$  \\ 
\\
HE0151$-$4326 & $\sim$11300& 2118 (E3 $\rightarrow$ E4) & $+2.3 $    & \ion{C}{4}  & \ion{C}{5}  $\rightarrow$ \ion{C}{4}   & $1.1\ten{48}$ & $\geq 3.90\ten{3}$ & $\leq 6.3$ \\
              & $\sim$8900 & 2118 (E3 $\rightarrow$ E4) & $+4.9 $    & \ion{Si}{4} & \ion{Si}{5} $\rightarrow$ \ion{Si}{4}  &               & $^h$               & $^h$        \\
              & $\sim$8900 & 2118 (E3 $\rightarrow$ E4) & $+12.2$    & \ion{C}{4}  & \ion{C}{5}  $\rightarrow$ \ion{C}{4}   &               & $\geq 3.90\ten{3}$ & $\leq 6.3$ \\
\\
Q1157+014     & $\sim$3000 & 2183 (E2 $\rightarrow$ E5) & $+2.4 $    & \ion{Si}{4} & \ion{Si}{5} $\rightarrow$ \ion{Si}{4}  & $2.9\ten{47}$ & $^g$               & $^g$        \\
\\
HE1341$-$1020 & $\sim$1300 & 2280 (E1 $\rightarrow$ E2) & $+6.8 $    & \ion{N}{5}  & \ion{N}{6}  $\rightarrow$ \ion{N}{5}   & $2.6\ten{47}$ & $\geq 1.75\ten{3}$ & $\leq 4.5$  \\
              &            &                            & $+10.8$    & \ion{C}{4}  & \ion{C}{5}  $\rightarrow$ \ion{C}{4}   &               & $\geq 3.00\ten{3}$ & $\leq 3.5$  \\  
\\
HS1603$+$3820 & $\sim$9500 &  471 (E1 $\rightarrow$ E2) & $+3.6 $    & \ion{C}{4}  & \ion{C}{5}  $\rightarrow$ \ion{C}{4}   & $1.9\ten{48}$ & $\geq 1.64\ten{4}$ & $\leq 4.0$  \\
\\
Q2343+125     & $\sim$24400& 1855 (E1 $\rightarrow$ E3) & $\leq 2.0$ & \ion{C}{4}  & \ion{C}{4}                             & $7.5\ten{47}$ & $^h$               & $^h$        \\
\enddata 
\tablenotetext{a}{Time interval in the observed frame in units of
  days. We list only the shortest time intervals on which variability 
  was detected since these yield the most stringent limit on the density
  of the absorbing gas (see \S\ref{sec:meth}).}
\tablenotetext{b}{Variation significance, defined as
  $S_\ew\equiv\dew/\sigma_\ew$; see equation~(\ref{eq:sew}). We have included 
  a low-significance measurement for Q2343+125 for completeness since we 
  have not detected more significant variability in this object.}
\tablenotetext{c}{Bolometric luminosity, calculated as $L_{\rm bol} =
  4.4\,\lambda L_{\lambda}(1450\,{\rm \AA})$.}
\tablenotetext{d}{Electron density, calculated by assuming that the
  variation time scale is the recombination time.}
\tablenotetext{e}{Absorber's distance from the flux source, calculated
  by assuming the quasi-static photoionization change.  These numbers
  become smaller by a factor of $\sim$4 if we use the SED of typical
  radio-loud quasar \citep{mat87} or high luminosity radio-quiet
  quasars \citep{dun10}.}
\tablenotetext{g}{No constraints can be placed because
  recombination processes to/from \ion{Si}{4} do not lead 
  to constraints on the density in straightforward way \citep{arn85}.}
\tablenotetext{h}{No constraints can be placed because of no time
  variation.}
\end{deluxetable}
\clearpage


\clearpage

\begin{figure}
 \begin{center}
  \includegraphics[height=4.5cm,angle=270]{figure2a.eps}
  \includegraphics[height=4.5cm,angle=270]{figure2b.eps}
  \includegraphics[height=4.5cm,angle=270]{figure2c.eps}
 \end{center}
\caption{Normalized spectra around the \lya, \ion{C}{4}, \ion{N}{5},
  and \ion{Si}{4} absorption lines of intrinsic systems at \voff\ =
  37,037 \kms\ (left), 4,129 \kms\ (middle), and 65,181 \kms\ (right) in
  the spectrum of HE0130$-$4021. The velocity scale is relative to the
  flux-weighted center of a line, calculated by equation~(4) of
  \citet{mis07a}. The 1st to 6th epoch spectra are shown with black,
  red, green, blue, light-blue, and purple histograms,
  respectively. Wavelength regions used for \ew\ measurements are
  marked with solid arrows. If lines are not detected with
  $>$5$\sigma$ level, upper limits of \ews\ are measured for the
  regions with dashed arrows. False patterns such as cosmic rays, bad
  columns, and order gaps, are marked with asterisks on the
  spectra.\label{fig:varplo}}
\end{figure}
\clearpage

\begin{figure}
 \begin{center}
  \includegraphics[height=4.4cm,angle=270]{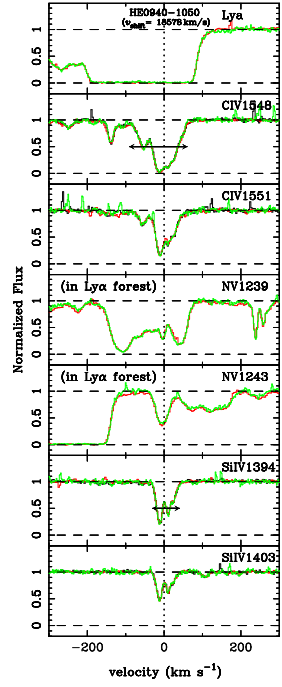}
 \end{center}
\caption{Same as Figure~\ref{fig:varplo}, but for the system at
  \voff\ = 18,578 \kms\ in the HE0940$-$1050 spectrum.}
\end{figure}

\begin{figure}
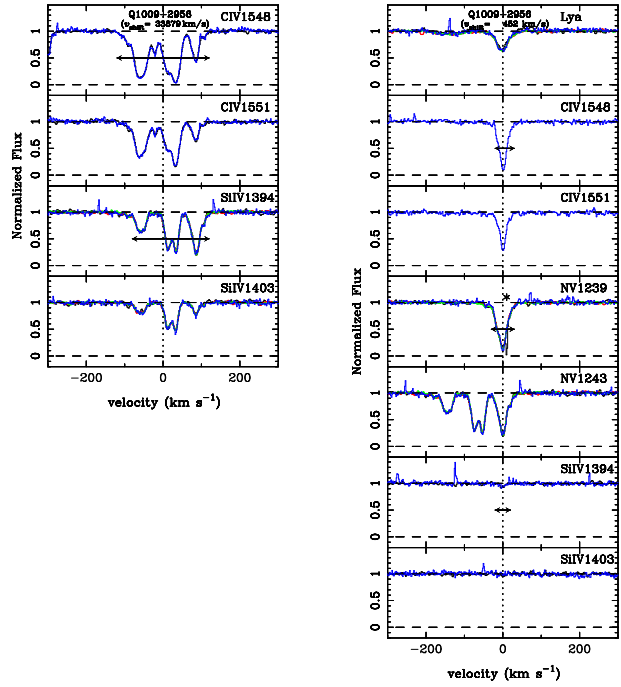

 \begin{center}
  \includegraphics[height=4.4cm,angle=270]{figure4a.eps}
  \includegraphics[height=4.4cm,angle=270]{figure4b.eps}
 \end{center}
\caption{Same as Figure~\ref{fig:varplo}, but for the systems at
  \voff\ = 33,879 \kms\ (left) and 452 \kms\ (right) in the
  Q1009$+$2,956 spectrum.}
\end{figure}
\clearpage

\begin{figure}
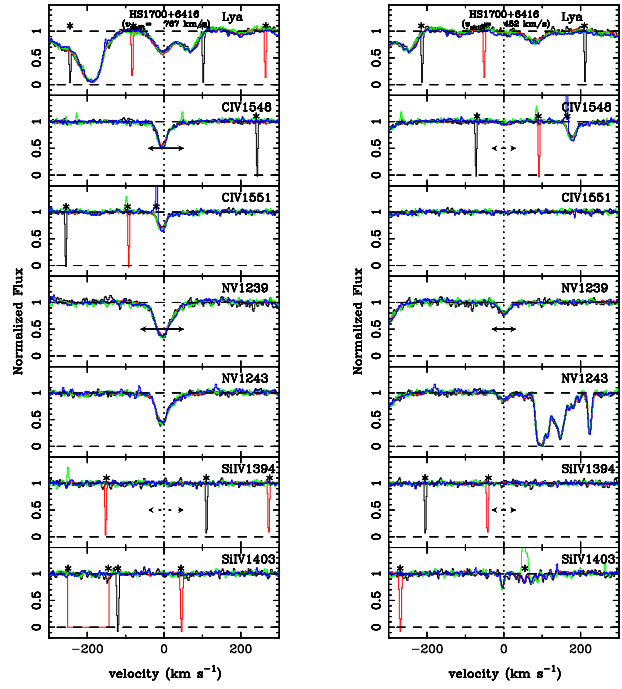

 \begin{center}
  \includegraphics[height=4.4cm,angle=270]{figure5a.eps}
  \includegraphics[height=4.4cm,angle=270]{figure5b.eps}
 \end{center}
\caption{Same as Figure~\ref{fig:varplo}, but for the systems at
  \voff\ = 767 \kms\ (left) and 452 \kms\ (right) in the HS1700$+$6416
  spectrum.}
\end{figure}

\begin{figure}
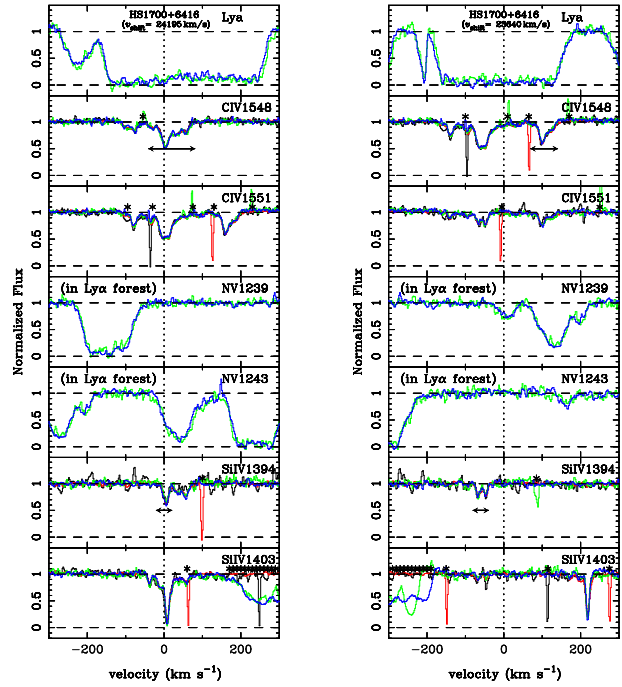

 \begin{center}
  \includegraphics[height=4.4cm,angle=270]{figure6a.eps}
  \includegraphics[height=4.4cm,angle=270]{figure6b.eps}
 \end{center}
\caption{Same as Figure~\ref{fig:varplo}, but for the systems at
  \voff\ = 24,195 \kms\ (left) and 23640 \kms\ (right) in the
  HS1700$+$6,416 spectrum.}
\end{figure}
\clearpage

\begin{figure}
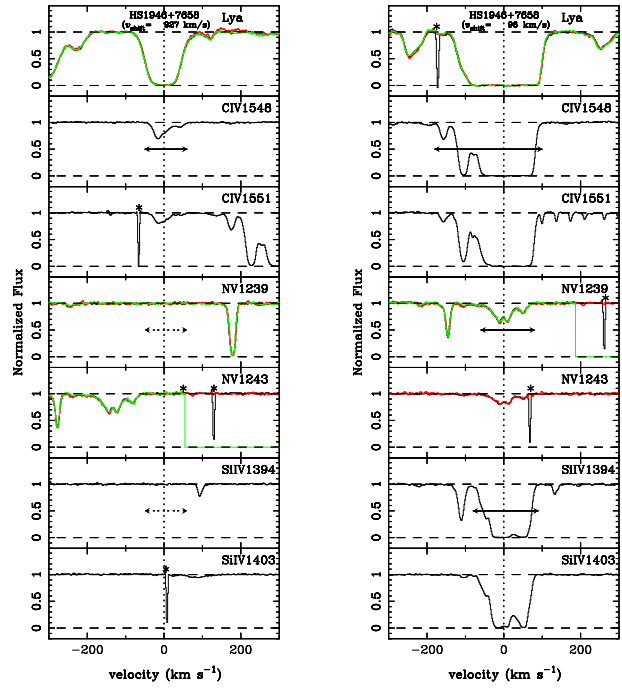

 \begin{center}
  \includegraphics[height=4.4cm,angle=270]{figure7a.eps}
  \includegraphics[height=4.4cm,angle=270]{figure7b.eps}
 \end{center}
\caption{Same as Figure~\ref{fig:varplo}, but for the systems at
  \voff\ = 927 \kms\ (left) and 96 \kms\ (right) in the HS1946$+$7658
  spectrum.}
\end{figure}

\begin{figure}
 \begin{center}
  \includegraphics[height=10cm,angle=0]{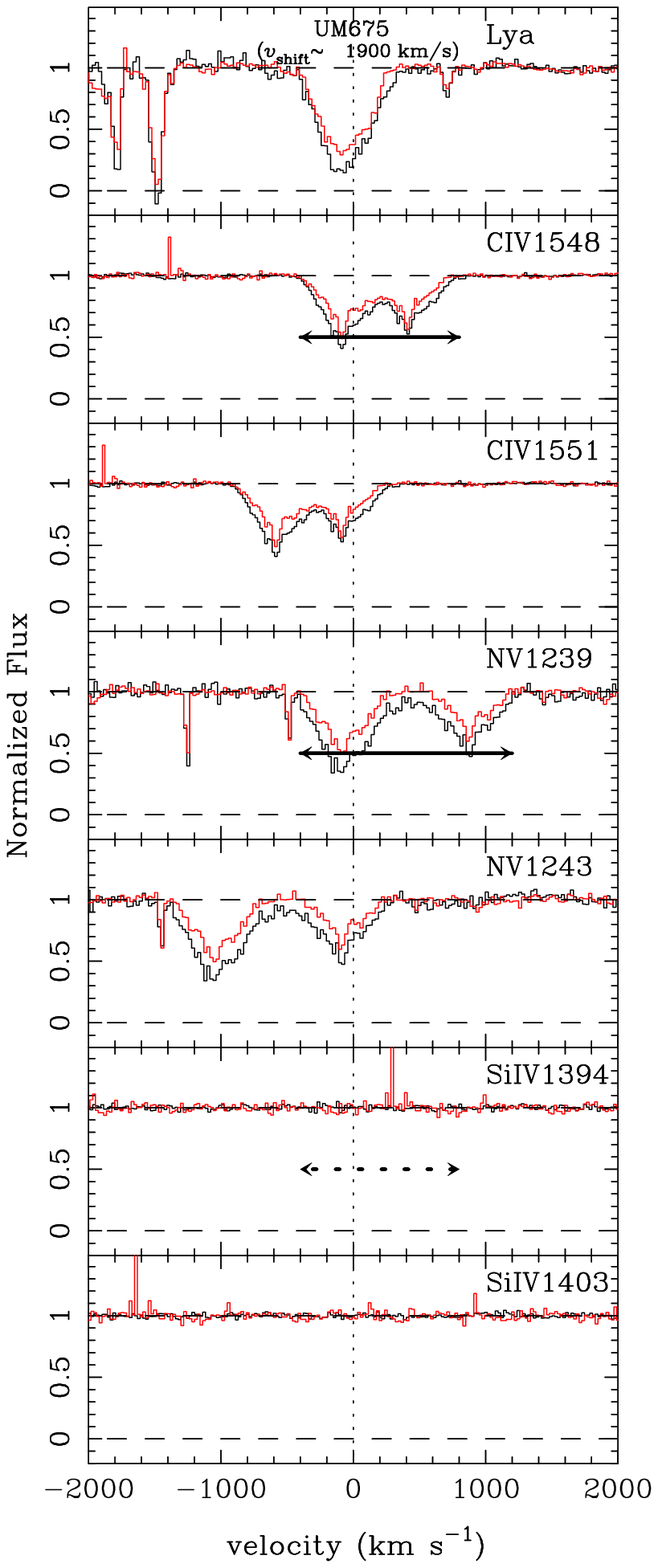}
 \end{center}
\caption{Same as Figure~\ref{fig:varplo}, but for the system at
  \voff\ $\sim$ 1,900 \kms\ in the UM675 spectrum.  The spectrum is
  resampled every 0.3\AA.\label{fig:varplo_UM675}}
\end{figure}
\clearpage

\begin{figure}
 \begin{center}
  \includegraphics[height=10cm,angle=0]{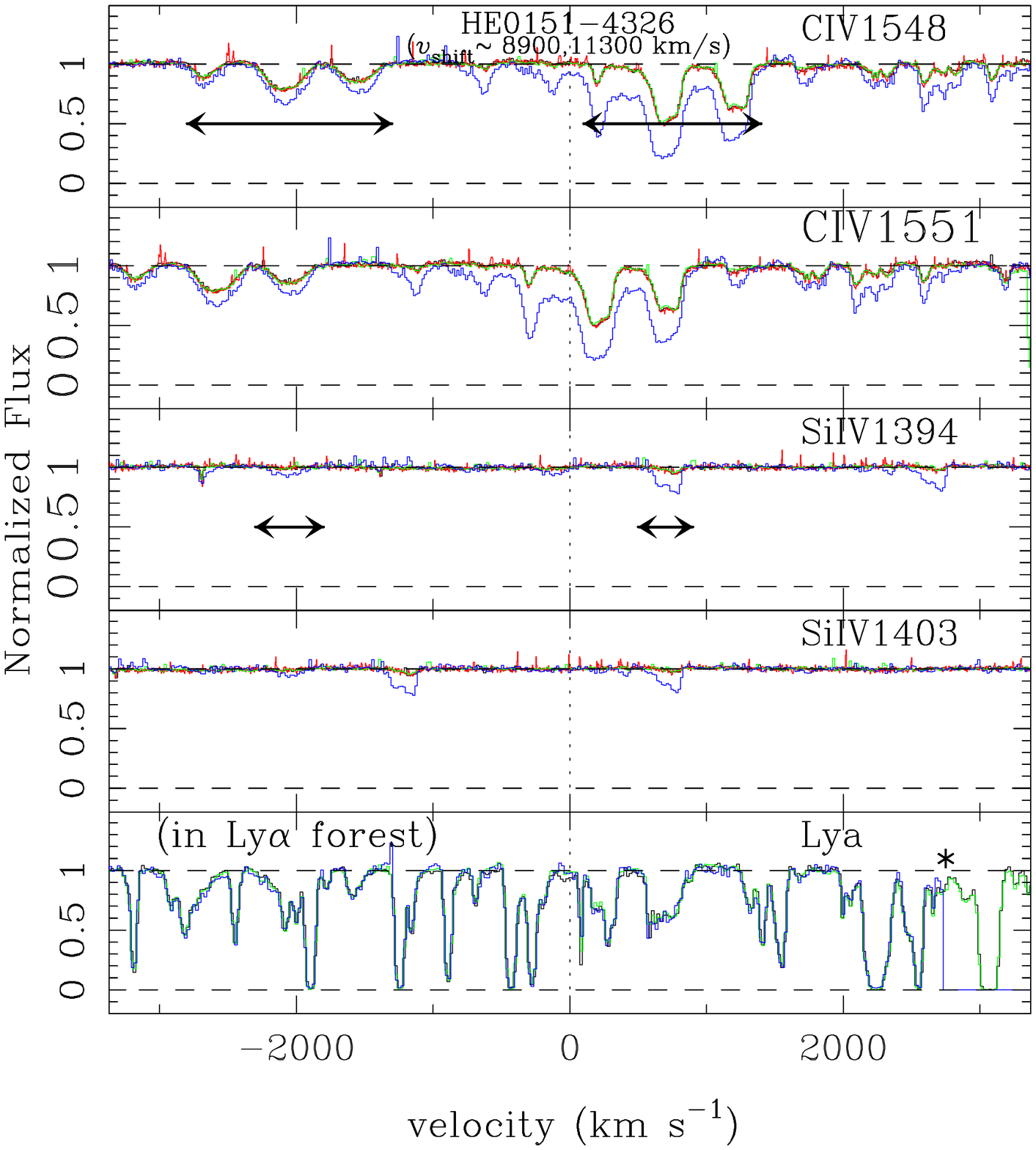}
 \end{center}
\caption{Same as Figure~\ref{fig:varplo}, but for the system at
  \voff\ $\sim$ 11,300 \kms\ and 8,900 \kms\ in the HE0151$-$4326
  spectrum.  The spectrum is resampled every
  0.3\AA.\label{fig:varplo_Q0151}}
\end{figure}

\begin{figure}
 \begin{center}
  \includegraphics[height=10cm,angle=0]{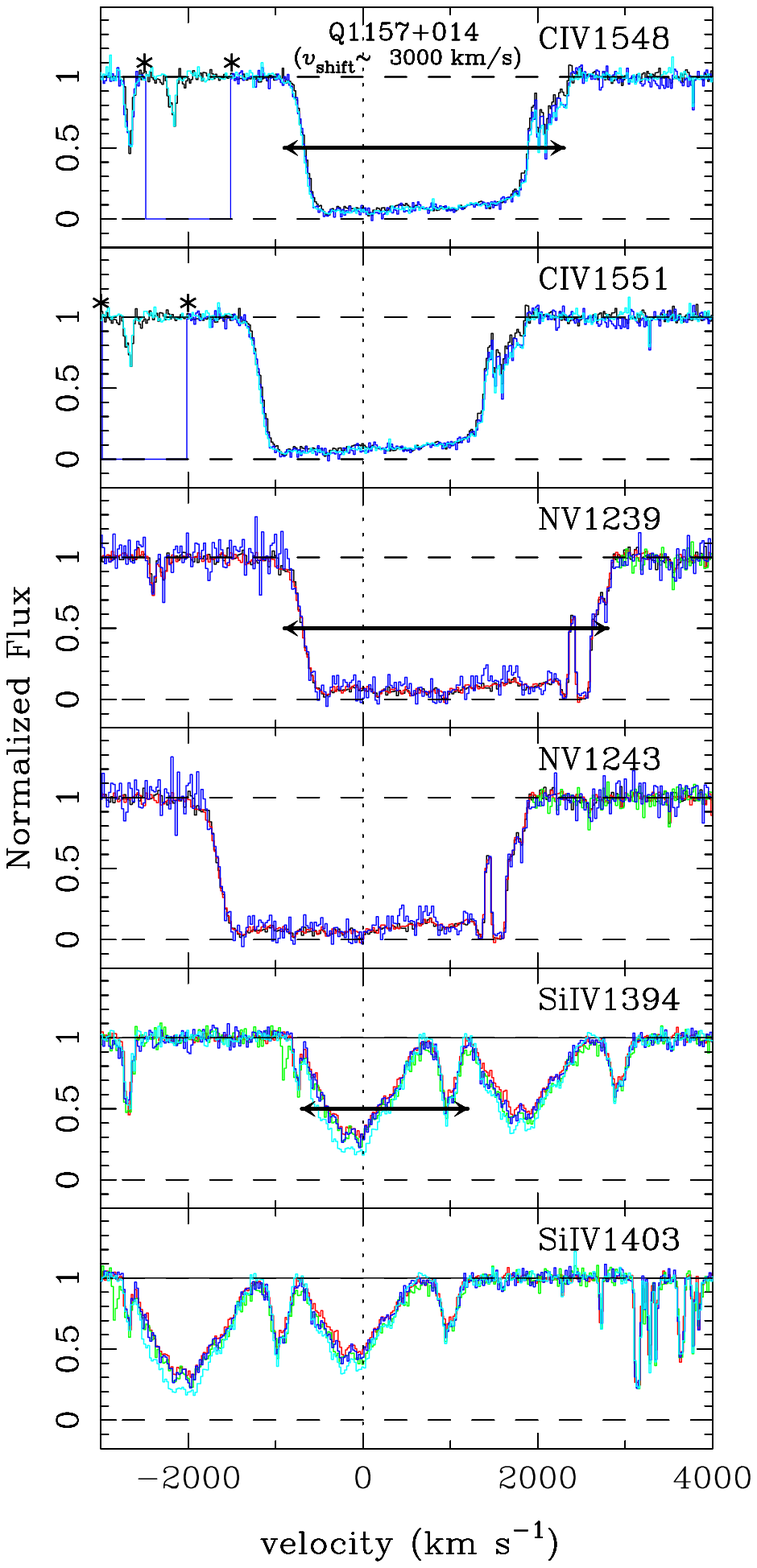}
 \end{center}
\caption{Same as Figure~\ref{fig:varplo}, but for the system in the
  \voff\ $\sim$ 3,000 \kms\ of Q1157$+$014 spectrum.  The spectrum is
  resampled every 0.3\AA.\label{fig:varplo_Q1157}}
\end{figure}
\clearpage

\begin{figure}
 \begin{center}
  \includegraphics[height=10cm,angle=0]{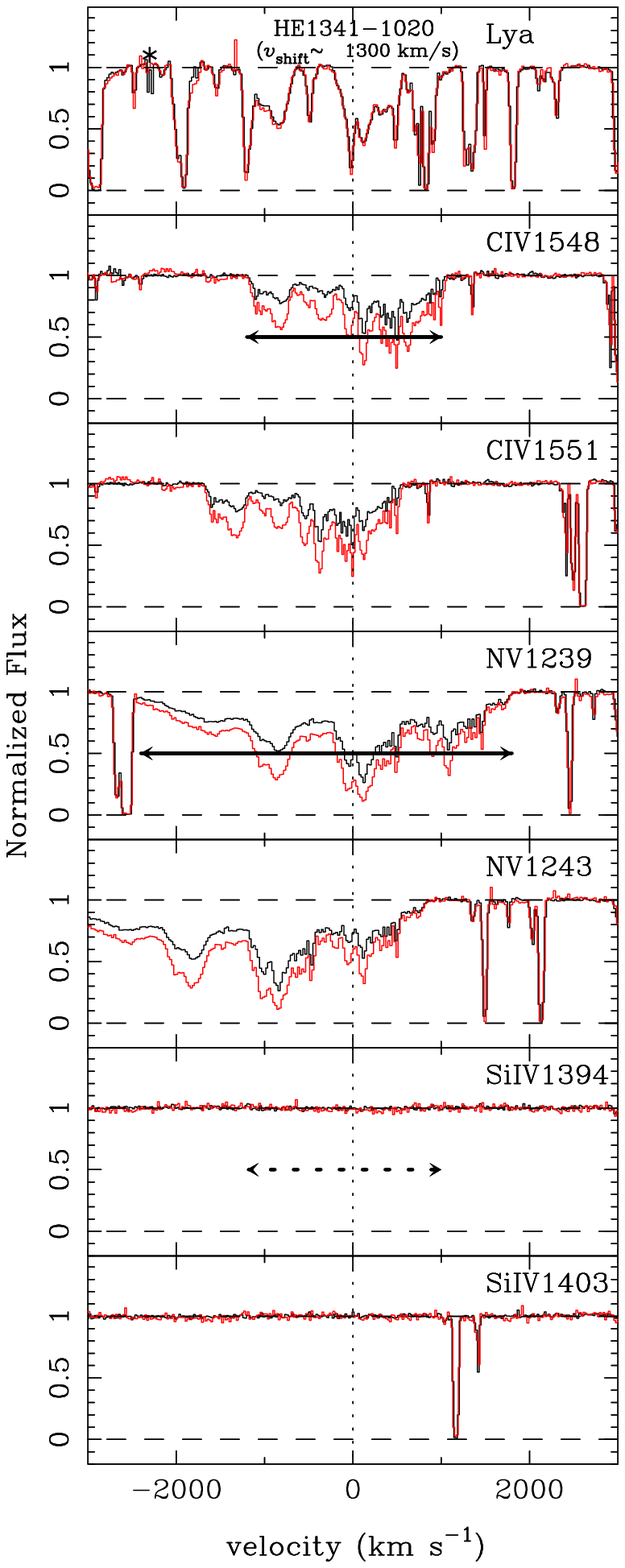}
 \end{center}
\caption{Same as Figure~\ref{fig:varplo}, but for the system in the
  \voff\ $\sim$ 1,300 \kms\ of HE1341$-$1020 spectrum.  The spectrum is
  resampled every 0.3\AA.\label{fig:varplo_Q1341}}
\end{figure}

\begin{figure}
 \begin{center}
  \includegraphics[height=10cm,angle=0]{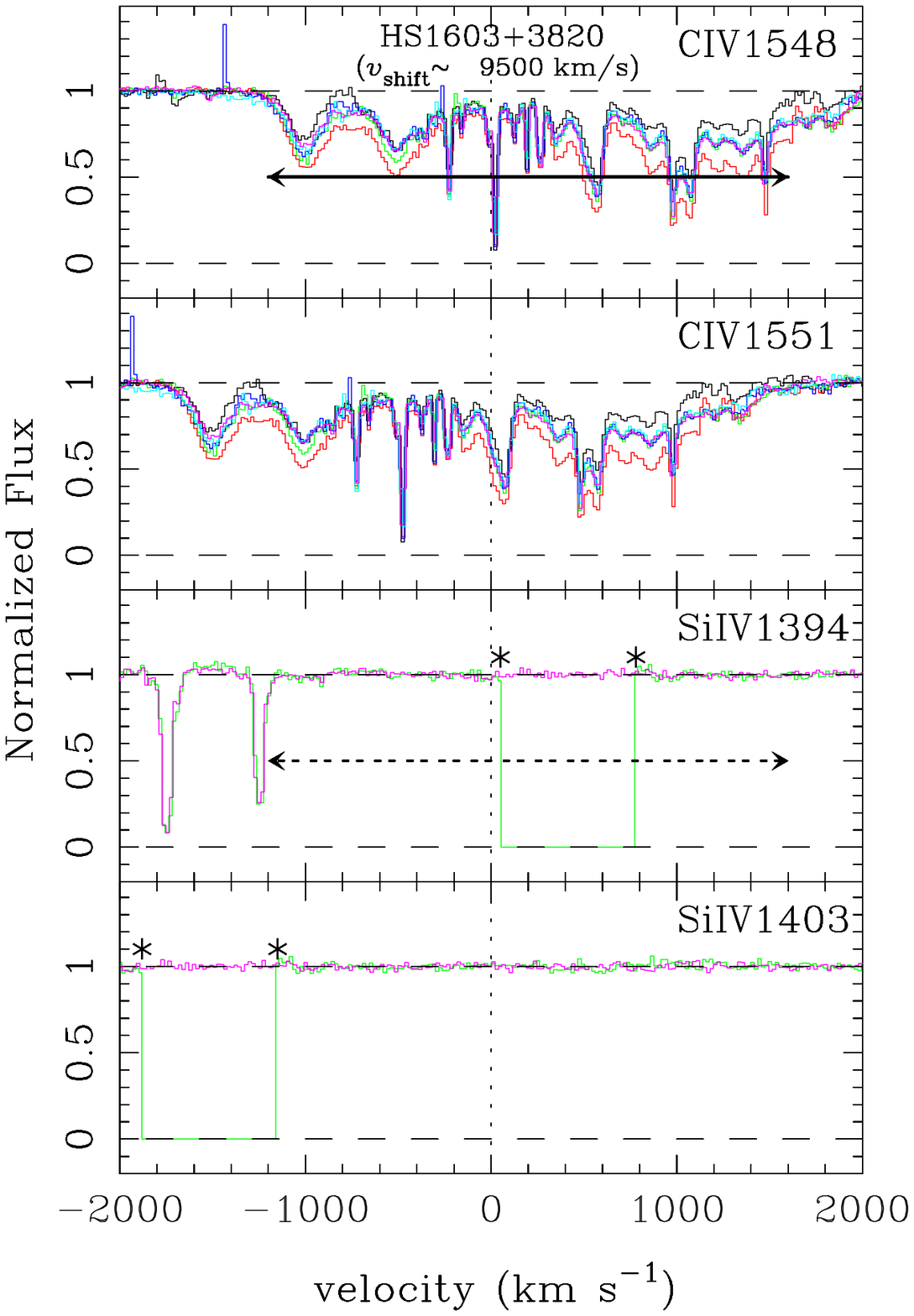}
 \end{center}
\caption{Same as Figure~\ref{fig:varplo}, but for the system at
  \voff\ $\sim$ 9,500 \kms\ in the HS1603+3820 spectrum.  The spectrum
  is resampled every 0.3\AA.\label{fig:varplo_HS1603}}
\end{figure}
\clearpage

\begin{figure}
 \begin{center}
  \includegraphics[height=10cm,angle=0]{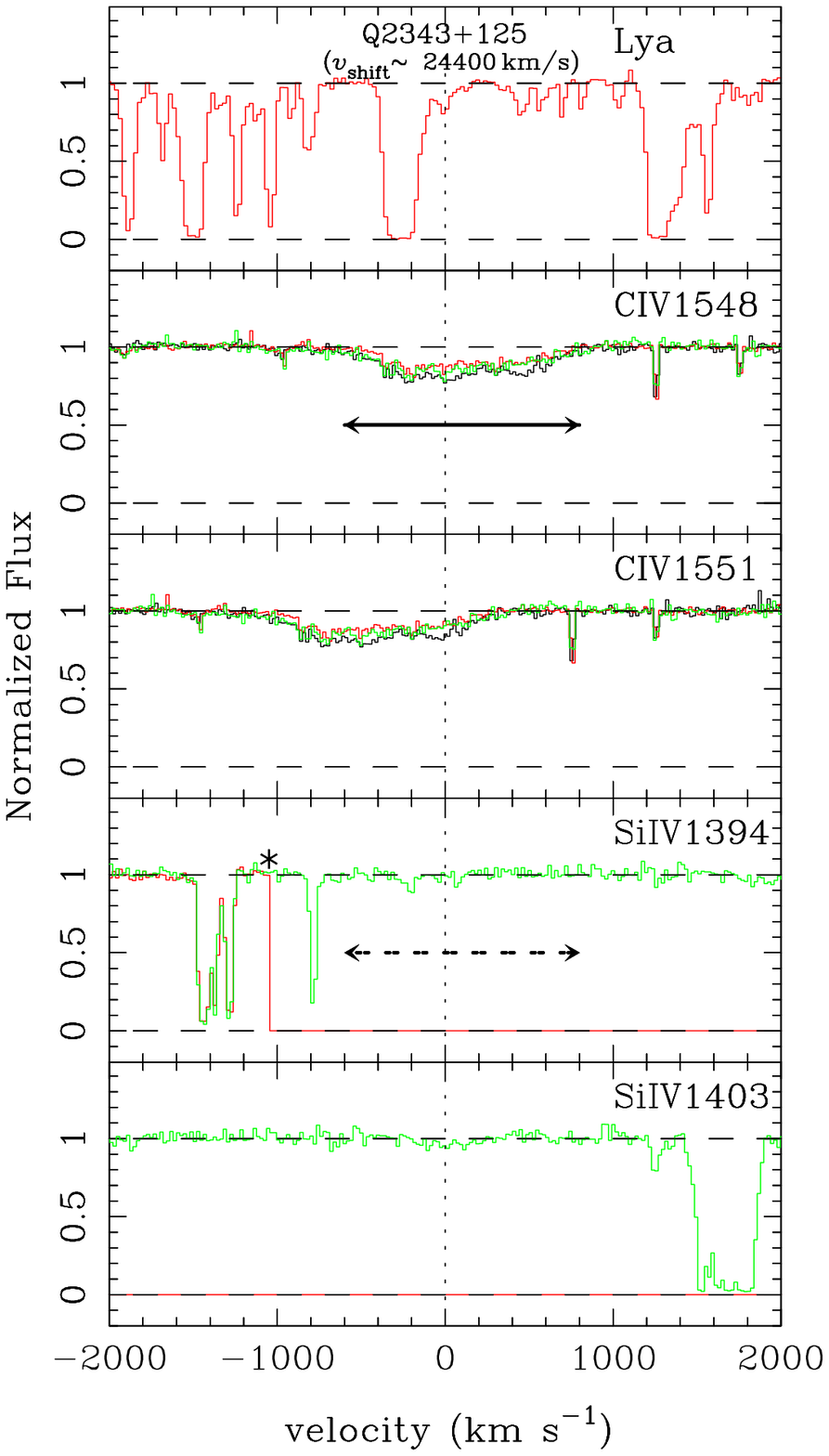}
 \end{center}
\caption{Same as Figure~\ref{fig:varplo}, but for the system at
  \voff\ $\sim$ 24,400 \kms\ in the Q2343$+$125 spectrum.  The spectrum
  is resampled every 0.3\AA.\label{fig:varplo_Q2343}}
\end{figure}

\clearpage

\clearpage

\clearpage

\clearpage

\clearpage

\clearpage

\clearpage

\clearpage

\clearpage

\end{document}